\documentclass[12pt,preprint]{aastex}
\begin{document}

\title{Accretion Disk Torqued by a Black Hole}

\author{Li-Xin Li}
\affil{Princeton University Observatory, Princeton, NJ 08544--1001, USA}
\email{E-mail: lxl@astro.princeton.edu}

\begin{abstract}
If a Kerr black hole is connected to a disk rotating around it by a magnetic 
field, the rotational energy of the Kerr black hole provides an energy source
for the radiation of the disk in addition to disk accretion. The black hole 
exerts a torque on the disk, which transfers energy and angular momentum between
the black hole and the disk. If the black hole rotates faster than the disk, 
energy and angular momentum are extracted from the black hole and transfered 
to the disk. The energy deposited into the disk is eventually radiated away by 
the disk, which will increase the efficiency of the disk. If the black hole 
rotates slower than the disk, energy and angular momentum are transfered from 
the disk to the black hole, which will lower the efficiency of the disk.

With suitable boundary conditions, quasi-steady state solutions are obtained for 
a thin Keplerian disk magnetically coupled to a Kerr black hole. By ``quasi-steady 
state'' we mean that any macroscopic quantity at a given radius in the disk slowly 
changes with time: the integrated change within one rotation period of the disk 
is much smaller than the quantity itself. We find that, the torque produced by the
magnetic coupling propagates only outward in the disk, the total radiation flux of 
the disk is a superposition of the radiation flux produced by the magnetic coupling 
and that produced by accretion. Most interestingly, a disk magnetically coupled to 
a rapidly rotating black hole can radiate without accretion. Such a disk has an 
infinite efficiency. For a specific example that the magnetic field touches the
disk at the inner boundary, the radial radiation profile is very different from 
that of a standard accretion disk: the emissivity index is significantly bigger,
most radiation comes from a region which is closer to the center of the disk.
The limitations of our model are briefly discussed, which include the assumption 
of a weak magnetic field, the ignorance of the instabilities of the disk and the 
magnetic field, and the ignorance of the radiation captured by the black hole
and the radiation returning to the disk.
\end{abstract}

\keywords{black hole physics --- accretion disks --- magnetic fields}

%\section 1
\section{Introduction}
In the standard theory of an accretion disk around a black hole it is assumed 
that there is no coupling between the disk and the central black hole
\citep{pri72,sha73,nov73,lyn74}. However, in the presence of a magnetic field, a
magnetic coupling between the disk and the black hole could exist and play an
important role in the balance and transportation of energy and angular momentum
(Zel'dovich and Schwartzman, quoted in Thorne 1974; Thorne, Price, and Macdonald 
1986; Blandford 1998, 1999; van Putten 1999; Gruzinov 1999; Li 2000c; Li and 
Paczy\'{n}ski 2000; Brown et al 2000). Though this issue was commented by 
Zel'dovich and Schwartzman more than twenty years ago and by many other people 
afterwards, a detailed and quantitative calculation has not appeared until 
\citet{li00b}. In the absence of the magnetic coupling, the energy source for the 
radiation of the disk is the gravitational energy of the disk (i.e., the 
gravitational binding energy between the disk and the black hole). But, if the 
magnetic coupling exists\footnote{In this paper we do not explore why a magnetic
connection between a black hole and a disk can exist or if such a magnetic connection
can exist. Instead, we assume a magnetic connection between a black hole and a disk
exists, and explore the consequences of the magnetic connection.}
and the black hole is rotating, the rotational energy of the black hole provides
an additional energy source for the radiation of the disk. With the magnetic coupling,
the black hole exerts a torque on the disk, which transfers energy and angular
momentum between the black hole and the disk. If the black hole
rotates faster than the disk, energy and angular momentum are extracted from the black
hole and transfered to the disk. The energy deposited into the disk is eventually
radiated away by the disk, which will increase the efficiency of the disk and
make the disk brighter than usual. If the black
hole rotates slower than the disk, energy and angular momentum are transfered from
the disk to the black hole, which will lower the efficiency of the disk and 
make the disk dimmer than usual. Therefore, the magnetic coupling between the black 
hole and the disk has important effects on the radiation properties of the disk. 
In this paper we investigate these effects in detail.
 
To be specific, we consider a model that a thin Keplerian disk rotates around a Kerr
black hole in the equatorial plane, and a magnetic field connects the black hole to 
the disk \citep{li00b}. This model is a variant of the standard 
Blandford-Znajek mechanism \citep{bla77,mac82,phi83a,phi83b,tho86}. In the
standard Blandford-Znajek mechanism, the magnetic field threading the black hole is 
assumed to close on a load which could be very far from the black hole. Due to
the large scale involved and the fact that the physics in the load region is ill 
understood, in some sense the Blandford-Znajek model is not well defined and thus the
involved physics is extremely complicated. Since the load is far from the black hole, 
the magnetic field suffers from the screw instability \citep{li00a}. Due to the lack 
of knowledge of the physics in the load region, we do not have a good
model for the load. The way people usually adopt is to assume the load is a resistor 
whose resistance is roughly equal to the resistance of the black hole 
\citep{mac82,phi83a,phi83b,tho86}. Since the load is so far from the black hole that 
the load cannot be casually connected to the black hole, it is hard to understand 
how the load can conspire with the black hole to have equal resistances and 
satisfy the impedance matching condition \citep{pun90}.

Compared with the standard Blandford-Znajek mechanism, the model we consider here 
is simpler and relatively well defined. In our model, the magnetic field lines 
threading the black hole close on the disk rather than a remote load. The disk is 
much better understood than the remote load though the magnetohydrodynamics (MHD) of 
the disk is still very complicated \citep{bal98,mil99,haw00}. In most cases the disk 
can be treated as fully ionized and thus its resistance is negligible compared with 
the resistance of the black hole which is several hundred Ohms. In other words, 
it is a good approximation to assume that the disk is perfectly conducting and the 
magnetic field lines are frozen in the disk. Certainly MHD instabilities have 
important effects on the dynamics of the disk and the magnetic field -- in fact 
people believe that the MHD instabilities play an important role in transporting 
angular momentum within the disk \citep{bal98}, but here we choose to ignore this 
topic since the problem of MHD instabilities is so complicated that a detailed 
discussion will be beyond the scope of the paper.

We assume that the disk is thin and Keplerian, lies in the equatorial plane of the 
black hole with the inner boundary being at the marginally stable orbit 
\citep{lyn69,bar70,nov73}. This requires that, the magnetic field is so weak that 
its influence on the dynamics of the particles in the disk is negligible. But, the 
role played a weak magnetic field in the balance and transportation of energy and 
angular momentum may be important. (In section \ref{sec5} we will justify the 
approximation of a weak magnetic field in detail.) With this ``weak magnetic field 
approximation'', the
particles in the disk move around the central black hole on circular orbits with a
superposition of a small radial inflow motion. If the inflow time-scale is much
longer than the dynamical (rotational) time-scale of the disk, the disk and the 
magnetic field can be in a quasi-steady state even though the magnetic field slowly
moves toward the central black hole along with the accretion. By ``quasi-steady 
state'' we mean that any macroscopic quantity at a given radius in the disk slowly 
changes with time: the integrated change within one rotational period of the
disk is much smaller than the quantity itself. With the assumption that 
the disk is thin and Keplerian, and the magnetic field and the disk are in a 
quasi-steady state, we will solve the conservation equations of energy and angular 
momentum, and calculate the radiation flux of the disk, the internal viscous torque
of the disk, and compare the results with those predicted by the standard theory
of an accretion disk.

Recently, several people have considered the magnetic coupling between the disk
and the material in the transition region between the disk and the black hole 
horizon, and have argued that such a coupling produces a torque at the inner
boundary of the disk \citep{kro99,gam99,ago00,haw00,kro01}. We emphasize that our 
model is
distinctly different from theirs: (1) In our model, the magnetic field connects
the disk to the black hole, thus the torque on the disk is produced by the black
hole; in their model, the magnetic field connects the disk to the material in
the transition region, thus the torque on the disk is produced by the material in
the transition region. (2) In our model, the magnetic field lines are assumed to
connect the disk surfaces to the black hole horizon at intermediate latitudes;
while in their model, the magnetic field is in the plane of the disk within the
inner edge of the disk thus magnetic reconnection can easily take place 
\citep{bla00}. (3) In our model, the magnetic field can touch the disk over a 
range of radii; while in their model, the magnetic field is attached to the disk 
only at the inner boundary of the disk \citep{ago00}. (4) In our model, we will 
show that the torque produced by the black hole propagates outward only and thus
the torque at the inner boundary of the disk is zero; in their model, the torque is
non-zero at the inner boundary of the disk and extends into the transition region
\citep{haw00,kro01}.
(5) In our model, energy and angular momentum can be extracted from the black hole 
even in the case with no accretion; in their model, in order to extract energy and 
angular momentum from the black hole accretion must exist so that negative energy
and angular momentum are carried into the black hole horizon by the accretion
material \citep{gam99}.

The paper is organized as follows: In section \ref{sec2} we discuss the transfer
of energy and angular momentum between the black hole and the disk arising from the
magnetic coupling between them. In section \ref{sec3} we solve the conservation 
equations of energy and angular momentum for a quasi-steady and thin Keplerian 
accretion disk torqued by a black hole and calculate the radiation flux, the internal 
viscous torque, and the total power of the disk. In
section \ref{sec3a} we argue that, with magnetic coupling to a rapidly
rotating black hole, a disk can radiate without accretion. Such a non-accretion disk
is purely powered by the rotational energy of the black hole and has an infinite
efficiency. In section \ref{sec4} we study a specific example: the magnetic field
lines touch the disk at a circle of a constant radius, and look for the consequences
of the magnetic coupling. We show that, the radial profile of the radiation flux
produced by the magnetic coupling is very different from that produced by accretion.
In section \ref{sec5}, we justify the assumptions made in the paper, and discuss the
limitations of our model which include the ignorance of the instabilities of the
disk and the magnetic field and the ignorance of the radiation captured by the black 
hole and the radiation returning to the disk. In section \ref{sec6} we draw our 
conclusions.

Throughout the paper we use the geometric units $G=c=1$ and the Boyer-Lindquist
coordinates $(t,r,\theta,\phi)$ \citep{mis73,wal84}.

%\section 2
\section{Transfer of Energy and Angular Momentum by the Magnetic Coupling
\label{sec2}}
In the presence of a magnetic field a black hole behaves like a conductor with
a surface resistivity $R_H = 4\pi \approx 377$ Ohms \citep{zna78,dam78,car79}.
So, when a black hole rotates in an external magnetic field, an electromotive
force (EMF) is induced on its horizon \citep{mac82,tho86}. As \citet{bla77} were 
first to note, the voltage drop along the magnetic field lines induced by the 
black hole rotation is huge enough to give rise a cascade production of 
electron-positron pairs, thereby producing a highly conducting
electron-positron pair plasma (magnetosphere or corona) around the black hole 
and the disk. Assume the magnetic field lines threading the black hole go through 
the magnetosphere and close on the disk which rotates around the black hole. Since 
the plasma disk is a good conductor, the rotation of the disk induces an EMF on 
the disk \citep{mac82,li00d}.

The black hole and the disk form a closed electric circuit, an electric current
flows along the magnetic field lines in the magnetosphere\footnote{It is assumed 
that in the magnetosphere the resistivity along the magnetic field lines is 
negligible, while the resistivity perpendicular to the magnetic field lines is 
large. Thus the electric current flows along the magnetic field lines in the 
magnetosphere without dissipation.} and close itself in the
disk and the black hole. Suppose the disk and the black hole rotates in the same
direction, then the black hole's EMF, ${\cal E}_H$, and the disk's EMF, 
${\cal E}_D$, have opposite signs. The direction of the electric current, and in 
turn the direction of the transfer of energy and angular momentum,
is determined by the sign of ${\cal E}_H + {\cal E}_D$. If ${\cal E}_H + 
{\cal E}_D > 0$, the black hole's EMF dominates the disk's EMF, so the black 
hole ``charges'' the disk: energy and angular momentum are transfered from the 
black hole to the disk. If ${\cal E}_H + {\cal E}_D < 0$, the disk's EMF 
dominates the black hole's EMF, so the disk ``charges'' the black hole, energy 
and angular momentum are transfered from the disk to the black hole. If 
${\cal E}_H + {\cal E}_D = 0$, the black hole's EMF balances the disk's EMF, 
then no energy and angular momentum are transfered between the black hole and 
the disk \citep{li00b}.

It is straightforward to calculate the electromagnetic power and torque on the
disk, using the standard electromagnetism. For the specific case that the magnetic
field lines touch the disk at a single radius, the calculations are carried 
out by \citet{li00b}. In this case, the sign of ${\cal E}_H + {\cal E}_D$ (and
thus the direction of the transfer of energy and angular momentum) is determined
by the sign of $\Omega_H-\Omega_D$, where $\Omega_H$ is the angular velocity of
the black hole which is constant over the black hole horizon, $\Omega_D$ is the
angular velocity of the disk at the radius where the magnetic field touches the
disk. If $\Omega_H > \Omega_D$, i.e. if the black hole rotates faster than the disk, 
energy and angular momentum are transfered from the black hole to the disk. If 
$\Omega_H <\Omega_D$, i.e. if the black hole rotates slower than the disk, energy 
and angular momentum are transfered from the disk to the black hole. If $\Omega_H  
=\Omega_D$, there is no transfer of energy and angular momentum between the 
black hole and the disk. For fixed values of the magnetic flux, the mass and the 
angular momentum of the black hole, and the resistance of the black hole, the 
power peaks at $\Omega_D = \Omega_H/2$ \citep{li00b}.

If the magnetic field is distributed over a differentially rotating disk, the 
formulae given in \citet{li00b} for the EMF of the disk, the power, and the torque
on the disk should be replaced with integrations over the radius of the disk. 
Assume the magnetic field is stationary and axisymmetric, and touches the disk at 
radii ranging from $r_1$ to $r_2$, then, the total EMF induced on the disk is
\begin{eqnarray}
    {\cal E}_D = {1\over 2\pi}\int_{r_1}^{r_2}\Omega_D\, {d\Psi_{HD}\over
    dr}\,dr\,,
    \label{demf1}
\end{eqnarray}
where $\Psi_{HD} = \Psi_{HD}(r)$ is the magnetic flux through a surface whose
boundary is a circle with a constant $r$ in the disk. In such a case, an infinite
number of adjacent infinitesimal poloidal electric current loops flow between
the black hole and the disk along the magnetic field lines connecting them. 
Each infinitesimal current loop produces an infinitesimal power and an 
infinitesimal torque on the disk, whose summation gives the the total power and
the total torque on the disk. Thus, assuming the disk is perfectly conducting, the
total power produced by the black hole on the disk is
\begin{eqnarray}
    P_{HD} = {1\over 4\pi^2} \int_{r_1}^{r_2} \left({d\Psi_{HD}\over d
    r}\right)^2\, {\Omega_D\left(\Omega_H - \Omega_D\right)\over - d Z_H/ dr}
    \,dr\,,
    \label{pow2}
\end{eqnarray}
where we have treated the black hole's resistance $Z_H$ as a function of the disk's
radius, which is defined by a map from the black hole horizon to the disk surface
given by the magnetic field lines. Similarly, the total torque produced by the black
hole on the disk is
\begin{eqnarray}
    T_{HD} = {1\over 4\pi^2} \int_{r_1}^{r_2} \left({d\Psi_{HD}\over d
    r}\right)^2\, {\Omega_H - \Omega_D\over - d Z_H/ dr}
    \,dr
    \label{toq2}\,.
\end{eqnarray}
From equation (\ref{pow2}) and equation (\ref{toq2}) we have $dP_{HD} = \Omega_D
dT_{HD}$, or
\begin{eqnarray}
    P_{HD} = \int_{r_1}^{r_2} \Omega_D {dT_{HD}\over dr} dr\,,
\end{eqnarray}
which follows from the assumption that the disk is perfectly conducting so the 
magnetic field is frozen in the disk.

For a Kerr black hole of mass $M_H$ and angular momentum $J_H = M_H a$, we have
\citep{tho86}
\begin{eqnarray}
    {d Z_H\over d\theta} = {R_H\over 2\pi}\,{r_H^2 + a^2 \cos^2\theta\over
         \left(r_H^2 + a^2\right) \sin\theta}\,,
\end{eqnarray}
where $r_H = M_H + \sqrt{M_H^2 - a^2}$ is the radius of the outer horizon of the 
black hole, $\theta$ is the polar angle coordinate on the horizon. Then, $dZ_H/
dr$ can be calculated through
\begin{eqnarray}
    {dZ_H\over dr} = {dZ_H\over d\theta}\,{d\theta\over dr}\,,
\end{eqnarray}
where $\theta = \theta(r)$ is a map between the $\theta$ coordinate on the horizon
and the $r$ coordinate on the disk, which is induced by the magnetic field lines
connecting the disk to the horizon. Since $d\theta/dr<0$ and $dZ_H/d\theta>0$,
we have $dZ_H/dr<0$.

%\section 3
\section{Quasi-Steady State for an Accretion Disk Torqued by a Black Hole
\label{sec3}}
Of particular interest is the case of an accretion disk in a steady state, which is 
simple enough for solving the equations of energy conservation and angular momentum
conservation and has been well studied when the magnetic coupling between the
black hole and disk does not exist \citep{pri72,sha73,nov73,lyn74,pag74,tho74}. In a 
steady state, the material of the disk advected with accretion toward smaller radii 
is balanced by the material advected from larger radii. Any macroscopic
(statistically averaged) quantity of the disk remains (approximately) unchanged for a
long period of time during accretion (say, over a time interval $>100$ rotation
periods of the disk). When there is a magnetic field connecting the black hole to the
disk, the disk and the magnetic field cannot be in an exactly steady state unless
there is no accretion at all. When there is accretion, the magnetic field frozen in
the disk slowly moves toward the central black hole as accretion goes on, which raises
the question how long a configuration of magnetic connection can be maintained. Due to
the complex topologies of the magnetic field, the magnetic field advected with
accretion toward smaller radii, which connects the black hole to the disk, does not
have to be balanced by the magnetic field advected from larger radii which also
connects the black hole to the disk. However, if the radial velocity of particles in
the disk is much smaller than their rotational velocity, the inflow time-scale is much
larger than the dynamical time-scale. Then, it is reasonable to assume that within one
rotation period, the global configuration of the magnetic field is almost unchanged,
and the overall change in a macroscopic quantity at a given
radius in the disk is much smaller than the quantity itself. When this condition is
satisfied, we say that the disk and the magnetic field are in a quasi-steady state. In
a quasi-steady state, a macroscopic quantity at a given radius may change significantly
over a long period of time and the magnetic connection between the black hole and the
disk may disappear at last, but they are approximately unchanged within one rotation
period of the disk. In this section we solve the equations for energy conservation and
angular momentum conservation for such a quasi-steady state disk magnetically coupled
to a black hole.

For a steady, axisymmetric, and thin Keplerian disk around a Kerr black hole, the
general relativistic equations of energy conservation and angular momentum conservation
have been investigated in detail by \citet{nov73,pag74}; and \citet{tho74}. Assume the
magnetic field is weak so that its influence on the dynamics of disk particles is
negligible, then a thin Keplerian disk is a good approximation\footnote{A magnetic 
field with negligible influence on the dynamics of the particles in the disk can play 
an important role in the balance and transportation of energy and angular momentum. In 
section \ref{sec5} we will justify the weak magnetic field approximation and the
importance of a weak magnetic field in the balance and transportation of energy and
angular momentum in detail.}. With the magnetic coupling between the black hole and the
disk being taken into account, for a quasi-steady state the conservation of angular 
momentum is described by
\begin{eqnarray}
    {d\over dr}\left(\dot{M}_D L^{+} -g\right) =
    4\pi r\left(FL^+ -H\right)\,,
    \label{con_ang}
\end{eqnarray}
where $\dot {M}_D \equiv dM_D/dt$ is the accretion rate of mass (measured by an observer 
at infinity; we use the convention $\dot {M}_D>0$ for accretion), $L^+$ is the specific 
angular momentum of a particle in the disk, $g$ is the internal viscous torque
of the disk, $F$ is the energy flux radiated away from the surface of the disk (measured
by an observer co-rotating with the disk), and $H$ is the flux of angular momentum 
transfered from the black hole to the disk by the magnetic field. The conservation of 
energy is described by
\begin{eqnarray}
    {d\over dr}\left(\dot{M}_D E^{+} -g\,\Omega_D\right) =
    4\pi r\left(FE^+ -H\Omega_D\right)\,,
    \label{con_ener}
\end{eqnarray}
where $E^+$ is the specific energy of a particle in the disk. When $H=0$, i.e., there 
is no magnetic coupling, equation (\ref{con_ang}) and equation (\ref{con_ener}) return 
to the equations derived by \citet{nov73}, and \citet{pag74}. $E^+$ is related to $L^+$ 
by \citep{pag74}
\begin{eqnarray}
    {dE^+\over dr} = \Omega_D {dL^+\over dr}\,.
    \label{el}
\end{eqnarray}
The flux of angular momentum transfered from the black hole to the disk, $H$, is 
defined by the torque $T_{HD}$ produced by the black hole on the disk
\begin{eqnarray}
    T_{HD} \equiv 2\pi \int_{r_1}^{r_2} H r dr\,.
    \label{flux_ang}
\end{eqnarray}
Comparing equation (\ref{flux_ang}) with equation (\ref{toq2}), we have
\begin{eqnarray}
    H = {1\over 8\pi^3} \left({d\Psi_{HD}\over d
    r}\right)^2\, {\Omega_H - \Omega_D\over -r d Z_H/ dr}\,,
    \label{toqh}
\end{eqnarray}
if the magnetic field is smoothly distributed on the disk.

In a quasi-steady state, $\dot{M}_D$ is constant throughout the disk (the
conservation of mass). Then, equation (\ref{con_ang}) and equation (\ref{con_ener}) 
can be solved for $F$ and $g$ by using equation (\ref{el}). The solutions are
\begin{eqnarray}
    F &=& {\dot{M}_D\over 4\pi r} f + {1\over 4\pi r}
    \left(-{d \Omega_D\over
    d r}\right)\left(E^+-\Omega_D L^+\right)^{-2}\times\nonumber\\
    &&\left[4\pi\int_{r_{ms}}^r\left(E^+-\Omega_D L^+\right)Hr dr + g_{ms}
    \left(E_{ms}^+ -\Omega_{ms} L_{ms}^+\right)\right]\,,
    \label{flux}
\end{eqnarray}
and
\begin{eqnarray}
    g = {E^+ - \Omega_D L^+\over -d \Omega_D/d r}\, 4\pi r F\,,
    \label{torque}
\end{eqnarray}
where the subscript ``$ms$'' denotes the marginally stable orbit which is assumed to 
be the inner boundary of a thin Keplerian disk, $g_{ms}$ is an integration constant, 
$f$ is defined by
\begin{eqnarray}
    f \equiv -{d \Omega_D\over d r} \left(E^+
    -\Omega_D L^+\right)^{-2}\int_{r_{ms}}^r\left(E^+-\Omega_D L^+\right)
    {dL^+\over dr}dr\,.
    \label{f0}
\end{eqnarray}
The integration of $f$ has been worked out by \citet{pag74} and is given by their
equation (15n), we do not display it here since it is lengthy. It is easy to check
that $f(r=r_{ms})=0$ and $f(r \gg r_{ms})\approx{3M_H\over 2r^2}$.

Suppose the magnetic field is distributed on the disk from $r=r_1$ to $r=r_2$, where
$r_2>r_1\ge r_{ms}$, then
\begin{eqnarray}
   \int_{r_{ms}}^r\left(E^+-\Omega_D L^+\right)Hr dr = \left\{
     \begin{array}{ll}
       0 & \mbox{if $r\le r_1$}\\
       \int_{r_1}^{r}\left(E^+-\Omega_D L^+\right)Hr dr & \mbox{if $r_1<r<r_2$}\\
       \int_{r_1}^{r_2}\left(E^+-\Omega_D L^+\right)Hr dr & \mbox{if $r\ge r_2$}
      \end{array}
   \right.\,.
   \label{int}
\end{eqnarray}
At the inner boundary of the disk, where $r=r_{ms}\le r_1$, we have $g =g_{ms}$. 
For a thin Keplerian disk, the ``no-torque inner boundary condition'' is
an excellent approximation \citep{nov73,muc82,abr89,pac00,arm00}. Though recently
this boundary condition has been challenged when there is a magnetic connection 
between the disk and the material in the transition region \citep{kro99,haw00}, in 
this paper we do not consider such a magnetic connection thus we adopt the 
``no-torque inner boundary condition''. The appearance of a magnetic coupling
between the black hole and the disk does not introduce a torque at the inner
boundary. This is because of that the integration in equation (\ref{int}) always 
vanishes at $r=r_{ms}$ for any $H$. In other words, the torque produced by the
magnetic coupling propagates outward only in the disk. Therefore, in the following 
discussion we take $g_{ms} = 0$. Then we have
\begin{eqnarray}
    F = {1\over 4\pi r}\left[\dot{M}_D f
    + 4\pi\left(-{d \Omega_D\over d r}\right)\left(E^+-\Omega_D L^+\right)^{-2}
    \int_{r_{ms}}^r\left(E^+-\Omega_D L^+\right)Hr dr\right]\,,
    \label{flux1}
\end{eqnarray}
and
\begin{eqnarray}
    g = {E^+ - \Omega_D L^+\over -d \Omega_D/d r}\,\dot{M}_D f
        + 4\pi \left(E^+ -\Omega_D L^+\right)^{-1} \int_{r_{ms}}^r
        \left(E^+ - \Omega_D L^+\right) Hr dr \,.
    \label{torque1}
\end{eqnarray}
Equation (\ref{flux1}) gives the radiation flux of the disk, equation (\ref{torque1})
gives the internal viscous torque of the disk. The first terms on the right hand
sides of equation (\ref{flux1}) and equation (\ref{torque1}) are the familiar results
for a standard accretion disk \citep{pag74}.

The integration of equation (\ref{con_ener}) gives the power (i.e., the luminosity) 
of the disk, which is the energy radiated by the disk per unit time as measured by 
an observer at infinity \citep{tho74}
\begin{eqnarray}
    {\cal L} &\equiv& 2\int_{r_{ms}}^\infty E^+ F 2\pi r dr \nonumber\\
      &=& \int_{r_{ms}}^\infty \left[{d\over d r}
          \left(\dot{M}_D E^+ -g\Omega_D\right) + 4\pi rH\Omega_D\right] dr
          \nonumber\\
      &=& \dot{M}_D \left(1-E_{ms}^+\right) + 4\pi \int_{r_{ms}}^\infty H
          \Omega_D r dr\,,
      \label{int_p}
\end{eqnarray}
where in the first line on the right hand side  the factor $2$ accounts for the
fact that a disk has two surfaces, in the third line we have used the boundary
conditions $E^+(r\rightarrow \infty) = 1$, $g\Omega_D (r\rightarrow \infty) = 0$, 
and $g\Omega_D (r=r_{ms}) = 0$. The power of the black hole, which is the energy 
transfered from the black hole to the disk per unit time as measured by an observer 
at infinity, is
\begin{eqnarray}
    {\cal L}_{HD} \equiv 2 P_{HD} = 4\pi\int_{r_{ms}}^\infty H\Omega_D r dr
    = 4\pi\int_{r_1}^{r_2} H\Omega_D r dr\,,
    \label{phd}
\end{eqnarray}
so equation (\ref{int_p}) can be written as
\begin{eqnarray}
    {\cal L} = \dot{M}_D \,\epsilon_0 + {\cal L}_{HD}\,,
    \label{bal}
\end{eqnarray}
where
\begin{eqnarray}
    \epsilon_0 = 1 - E_{ms}^+
    \label{eff}
\end{eqnarray}
is the efficiency of accretion, i.e., the efficiency of the disk when the magnetic 
coupling between the black hole and the disk does not exist. For a Schwarzschild 
black hole, $\epsilon_0 \approx 0.06$; for a Kerr black hole of $s \equiv a/M_H = 
0.998$, $\epsilon_0 \approx 0.32$; for an extreme Kerr black hole of $s = 1$, 
$\epsilon_0\approx 0.42$ \citep{tho74}. For any thin Keplerian disk around a Kerr 
black hole, $\epsilon_0 < 0.42$ always.

Equation (\ref{bal}) describes the global balance of energy for a quasi-steady
accretion disk magnetically coupled to a Kerr black hole. ${\cal L}$ is the total
power of the disk. $\dot{M}_D\,\epsilon_0$ represents the rate of change in the
gravitational energy of the disk, as in the standard theory of an accretion disk.
${\cal L}_{HD}$ represents the rate of energy transfered from the black hole to the
disk, which is absent in the standard theory of an accretion disk.

From equation (\ref{bal}), the total efficiency of the disk is
\begin{eqnarray}
    \epsilon \equiv {{\cal L}\over \dot{M}_D} = \epsilon_0 +
    {{\cal L}_{HD}\over \dot{M}_D}\,.
    \label{eff2}
\end{eqnarray}
If ${\cal L}_{HD} = 0$, i.e. there is no energy transfer between the black hole 
and the disk by the magnetic coupling -- as the case in the standard theory of an 
accretion disk, $\epsilon = \epsilon_0$ and the power of the disk purely comes 
from the gravitational energy of the disk. If ${\cal L}_{HD}>0$, which is the case 
when the black hole rotates faster than the disk, energy is transfered from the 
black hole to the disk. Then there are two sources for the power of the disk: one 
is the gravitational energy of the disk (represented by $\dot{M}_D\,\epsilon_0$), 
the other is the rotational energy of the black hole (represented by 
${\cal L}_{HD}$). The efficiency of the disk is increased by the magnetic coupling: 
$\epsilon > \epsilon_0$. For very small $\dot{M}_D$, an efficiency $\epsilon\gg 1$ 
can be achieved. If there is no accretion at all, the total efficiency of the disk 
is infinite. If ${\cal L}_{HD}<0$, which is the case when the black hole rotates 
slower than the disk, energy is transfered from the disk to the black hole. Then, 
the efficiency of the disk is decreased by the magnetic coupling: $\epsilon <
\epsilon_0$. Since ${\cal L}$ cannot be negative, we must have
\begin{eqnarray}
    \dot{M}_D\epsilon_0 \ge - {\cal L}_{HD}\,.
\end{eqnarray}

%\section 3a
\section{Radiation without Accretion: A Disk Powered by a Black Hole
\label{sec3a}}
For a disk magnetically coupled to a rapidly rotating black hole, an extremely 
interesting feature is that the disk can radiate without accretion. This can be seen
directly from equation (\ref{flux1}) and equation (\ref{int_p}): when $\dot{M}_D = 
0$, $F$ is non-negative over the disk and ${\cal L}$ is positive if
\begin{eqnarray}
    \int_{r_{ms}}^r \left(E^+ - \Omega_D L^+\right) Hr dr \ge 0
    \label{cond_x}
\end{eqnarray}
for any $r>r_{ms}$, and
\begin{eqnarray}
    \int_{r_{ms}}^{\infty} \left(E^+ - \Omega_D L^+\right) Hr dr > 0\,.
    \label{cond_y}
\end{eqnarray}
From equation (\ref{toqh}), the sign of $H$ is determined by the sign of $\Omega_H
-\Omega_D$ since $dZ_H/dr < 0$. Since $\Omega_H$ is constant over the horizon of 
the black hole and $d\Omega_D/dr <0$ for $r>r_{ms}$ over the disk, conditions
(\ref{cond_x}) and (\ref{cond_y}) are equivalent to the requirement that $H\ge 0$ 
over the disk and $H>0$ at least over an interval of $r>r_{ms}$.

For such a non-accretion disk, the radiation flux and the internal viscous torque
are respectively
\begin{eqnarray}
    F = {1\over r}\left(-{d \Omega_D\over d r}\right)
    \left(E^+-\Omega_D L^+\right)^{-2}
    \int_{r_{ms}}^r\left(E^+-\Omega_D L^+\right)Hr dr\,,
    \label{flux2}
\end{eqnarray}
and
\begin{eqnarray}
    g = 4\pi \left(E^+ -\Omega_D L^+\right)^{-1} \int_{r_{ms}}^r
        \left(E^+ - \Omega_D L^+\right) Hr dr \,,
    \label{torque2}
\end{eqnarray}
where $H$ is given by equation (\ref{toqh}). The power of the disk comes purely from 
the rotational energy of the black hole
\begin{eqnarray}
    {\cal L} = {\cal L}_{HD} = 4\pi \int_{r_{ms}}^\infty H\Omega_D r dr\,.
    \label{purl}
\end{eqnarray}
Such a disk is powered by a black hole and has an infinite efficiency.

We emphasize that ``radiation without accretion'' is a specific feature of our model: 
a disk coupled to a black hole with a magnetic field. A standard accretion disk
radiates only if the accretion rate is non-zero, and the efficiency of a standard
accretion disk is always smaller than $0.42$. (The maximum efficiency $0.42$ can be 
reached only for a disk around an extreme Kerr black hole with $a = M_H$.) For the 
model of a disk magnetically coupled to the material in the transition region, though 
it has been demonstrated that the efficiency of the disk is unbounded from above if 
the black hole rotates faster than the disk \citep{ago00}, a state with a zero 
accretion rate and a finite power can never be realized since in that model in order 
to extract energy from the black hole material with negative energy must fall into 
the black hole thus accretion must exist \citep{gam99}.

%\section 4
\section{An Example: the Magnetic Field Touches the Disk at a Circle
\label{sec4}}
Suppose the magnetic field touches the disk at a circle with a radius $r=r_0 >
r_{ms}$, i.e.
\begin{eqnarray}
    H = A_0\, \delta (r-r_0)\,,
    \label{hdel}
\end{eqnarray}
where $A_0$ is a constant and $\delta (x)$ is a Dirac $\delta$-function which
satisfies
\begin{eqnarray}
    \delta(x) = 0\,,
    \label{del1}
\end{eqnarray}
for any $x\neq 0$, and
\begin{eqnarray}
    \int_{-\infty}^{\infty}y(x)\delta(x) dx = y(0)\,,
    \label{del2}
\end{eqnarray}
for any smooth function $y(x)$. Though this is a simple and highly ideal
case, it is fundamental in understanding the effect of the magnetic coupling on
the energetic process of the disk. For any given distribution of a magnetic field
on the disk, the corresponding $H(r)$ can always be written as
\begin{eqnarray}
    H(r) = \int_{r_{ms}}^\infty H(r^\prime) \delta (r^\prime - r) dr^\prime\,.
    \label{hr}
\end{eqnarray}
Since the energy flux and the internal torque given by equation (\ref{flux1})
and equation (\ref{torque1}) are linear functionals of $H(r)$, the results for
any given distribution of a magnetic field can be obtained from the results for 
the simple in equation (\ref{hdel}) by linear superpositions.
Furthermore, for a realistic case of a smooth distribution of a magnetic field
with $r_b-r_a\ll r_a$, where $r_a$ and $r_b$ are the radii of the boundaries of
an annular region in the disk within which most of the magnetic field lines are
concentrated, equation (\ref{hdel}) is a good approximation for $H$ if we take
$r_0 \approx (r_a+r_b)/2$ and
\begin{eqnarray}
    A_0 \approx {2\over r_a + r_b} \int_{r_a}^{r_b} H r dr\,.
\end{eqnarray}
Thus, the simple case given by equation (\ref{hdel}) is meaningful not only in
mathematics but also in practice.

Inserting equation (\ref{hdel}) into equation (\ref{flux_ang}) and comparing the
result with equation (3) of \citet{li00b}, we obtain
\begin{eqnarray}
    A_0 = {1\over 2\pi r_0} \left({\Delta\Psi_{HD}\over 2\pi}\right)^2\,
        {\Omega_H-\Omega_0 \over Z_H}\,,
    \label{h0}
\end{eqnarray}
where $\Omega_0 \equiv \Omega_D (r_0)$, $\Delta\Psi_{HD}$ is the magnetic flux
connecting the disk to the black hole. The sign of $A_0$ is determined by the sign
of $\Omega_H - \Omega_0$: $A_0 >0$ if $\Omega_H > \Omega_0$; $A_0 <0$ if $\Omega_H
< \Omega_0$. For a thin Keplerian disk around a Kerr black hole, The ratio of 
$\Omega_0/\Omega_H$ is
\begin{eqnarray}
    {\Omega_0\over\Omega_H} = {2\omega\over s} \left[1+(1-s^2)^{1/2}\right]\,,
    \label{wdh}
\end{eqnarray}
where $s\equiv a/M_H$, $\omega\equiv M_H\Omega_0$ is a function of $s$ and 
$r_0/M_H$
\begin{eqnarray}
    \omega ={1\over s+ \left({r_0\over M_H}\right)^{3/2}}\,.
    \label{omega}
\end{eqnarray}
The critical case of $\Omega_H = \Omega_0$ (i.e. $A_0 = 0$) is determined by
\begin{eqnarray}
    \omega \left(s, {r_0\over M_H}\right) =
    {s\over 2 \left[1+(1-s^2)^{1/2}\right]}\,.
    \label{crit}
\end{eqnarray}
For any given value of $r_0/M_H$, or equivalently, for any given value of
$r_0/r_{ms}$, equation (\ref{crit}) can be solved for $s$ to obtain the critical 
spin $s_c$. $\Omega_H > \Omega_0$ (thus $A_0>0$) for $s>s_c$, $\Omega_H < \Omega_0$
(thus $A_0<0$) for $s<s_c$. It is easy to show that $s_c$ is a monotonically
decreasing function of $r_0/r_{ms}$. For $r_0/r_{ms} = 1$, $s_c \approx 
0.3594$~\footnote{Another solution to equation (\ref{crit}) for $r_0/r_{ms}=1$ is 
$s = 1$, but here we do not consider the case of an extreme Kerr black hole. In 
other words, we restrict ourselves to the cases with $0\le s<1$.}; as $r_0/r_{ms}$ 
increases, $s_c$ decreases quickly. As $r_0/r_{ms}\rightarrow\infty$, $s_c 
\rightarrow 0$.

\begin{figure}
\epsscale{0.9}
\plotone{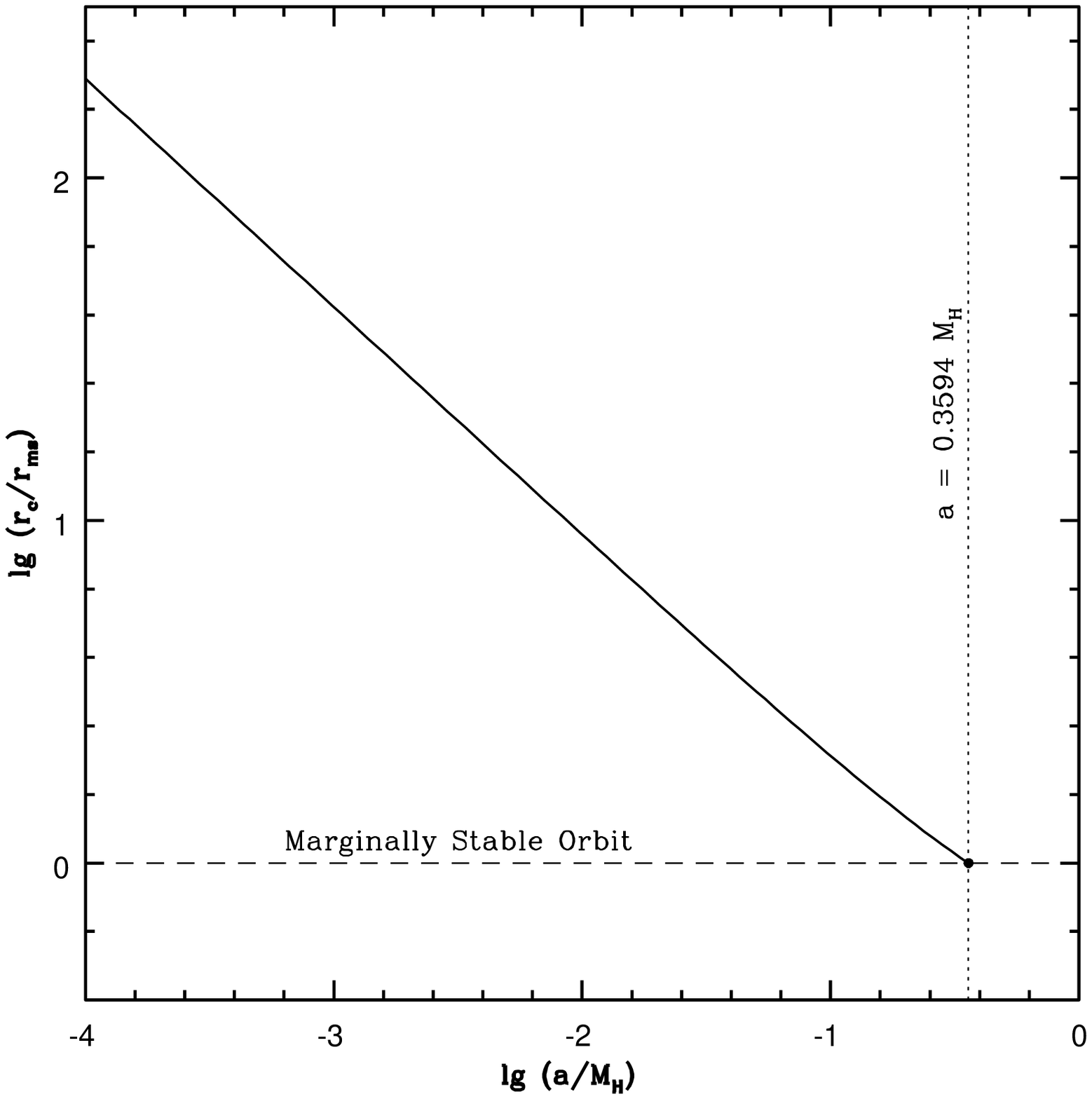}
\caption{The ratio of the co-rotation radius to the marginally stable radius of a 
thin Keplerian disk, $r_c/r_{ms}$, as a function of the black hole spin, $s=a/M_H$. 
When $0< s< 0.3594$, the co-rotation radius is given by equation (\ref{coro}), 
which is shown with the solid curve. Beyond the co-rotation radius the black hole 
rotates faster than the disk, within the co-rotation radius the black hole rotates 
slower than the disk. When $s > 0.3594$, the co-rotation radius does not exist: 
the black hole always rotates faster than the disk. [If the label of the horizontal 
axis is replaced with $\lg s_c$ where $s_c$ is the critical spin of the black hole 
defined by equation (\ref{crit}), the label of the vertical axis is replaced with 
$\lg \left(r_0/r_{ms}\right)$ where $r_0$ is the radius where the magnetic field 
touches the disk, then the same solid curve gives the relation $s_c = s_c
\left(r_0\right)$.]
\label{fig1}}
\end{figure}

When $0.3594<s<1$, the black hole rotates faster than a particle in the disk at any
radius ($\ge r_{ms}$). When $0 < s<0.3594$, there exists a co-rotation radius in the
disk defined by equation (\ref{crit}), beyond which the black hole rotates faster 
than the disk, within which
the black hole rotates slower than the disk. When the magnetic field lines touch the
disk beyond the co-rotation radius, energy and angular momentum are transfered from
the black hole to the disk. When the magnetic field lines touch the disk inside the
co-rotation radius, energy and angular momentum are transfered from the disk to the
black hole. The co-rotation radius, which we denote by $r_c$, can be solved from
equation (\ref{crit})
\begin{eqnarray}
    r_c = M_H \left\{{2\over s}\left[1+\left(1-s^2\right)^{1/2}\right] -
    s\right\}^{2/3}\,, \hspace{1cm} 0 <s< 0.3594\,.
    \label{coro}
\end{eqnarray}
The ratio $r_c/r_{ms} = (r_c/M_H) (r_{ms}/M_H)^{-1}$ as a function of $s$ is plotted
in Fig \ref{fig1}, from which we see that $r_c/r_{ms}$ is a monotonically decreasing
function of $s$: $r_c \rightarrow r_{ms}$ as $s \rightarrow 0.3594$, $r_c\rightarrow
\infty$ as $s \rightarrow 0$. The co-rotation radius does not exist when $s>0.3594$
for which the black hole always rotates faster than the disk. [Fig. \ref{fig1} also
shows $s_c = s_c \left(r_0\right)$ if we replace the label of the horizontal axis
with $\lg s_c$ and replace the label of the vertical axis with $\lg\left(r_0/
r_{ms}\right)$.]

Inserting equation (\ref{hdel}) into equation (\ref{flux1}) and equation 
(\ref{torque1}), we obtain
\begin{eqnarray}
    F = \left\{\begin{array}{ll}
       {1\over 4\pi r} \dot{M}_D f\,, & r_{ms}\le r< r_0 \\
       {1\over 4\pi r} \left[\dot{M}_D f + 4\pi r_0 A_0 \left(E_0^+ -\Omega_0 
       L_0^+\right) \left(-{d\Omega_D \over dr}\right)\left(E^+ -\Omega_D 
       L^+\right)^{-2} \right]\,, & r>r_0
       \end{array}
       \right.\,,
       \label{fdel}
\end{eqnarray}
and
\begin{eqnarray}
    g = \left\{\begin{array}{ll}
       {E^+ - \Omega_D L^+\over -d \Omega_D/d r}\,\dot{M}_D f\,, &
       r_{ms}\le r< r_0\\
       {E^+ - \Omega_D L^+\over -d \Omega_D/d r}\,\dot{M}_D f +
       4\pi r_0 A_0 \left(E_0^+ -\Omega_0
       L_0^+\right) \left(E^+ -\Omega_D L^+\right)^{-1}\,,
       & r>r_0
       \end{array}
       \right.\,.
       \label{gdel}
\end{eqnarray}
The expressions for $E^+ -\Omega_D L^+$, $d\Omega_D/dr$, and $f$ can be found in
\citet{pag74}. Equation (\ref{fdel}) and equation (\ref{gdel}) clearly show that
the torque produced by the magnetic coupling at $r=r_0$ propagates outward only,
thus the magnetic coupling between the black hole and disk has effects only in 
the region beyond the circle $r=r_0$ in the disk.

%\section 4.1
\subsection{$\dot{M}_D = 0$
\label{sec4.1}}
\begin{figure}
\epsscale{0.9}
\plotone{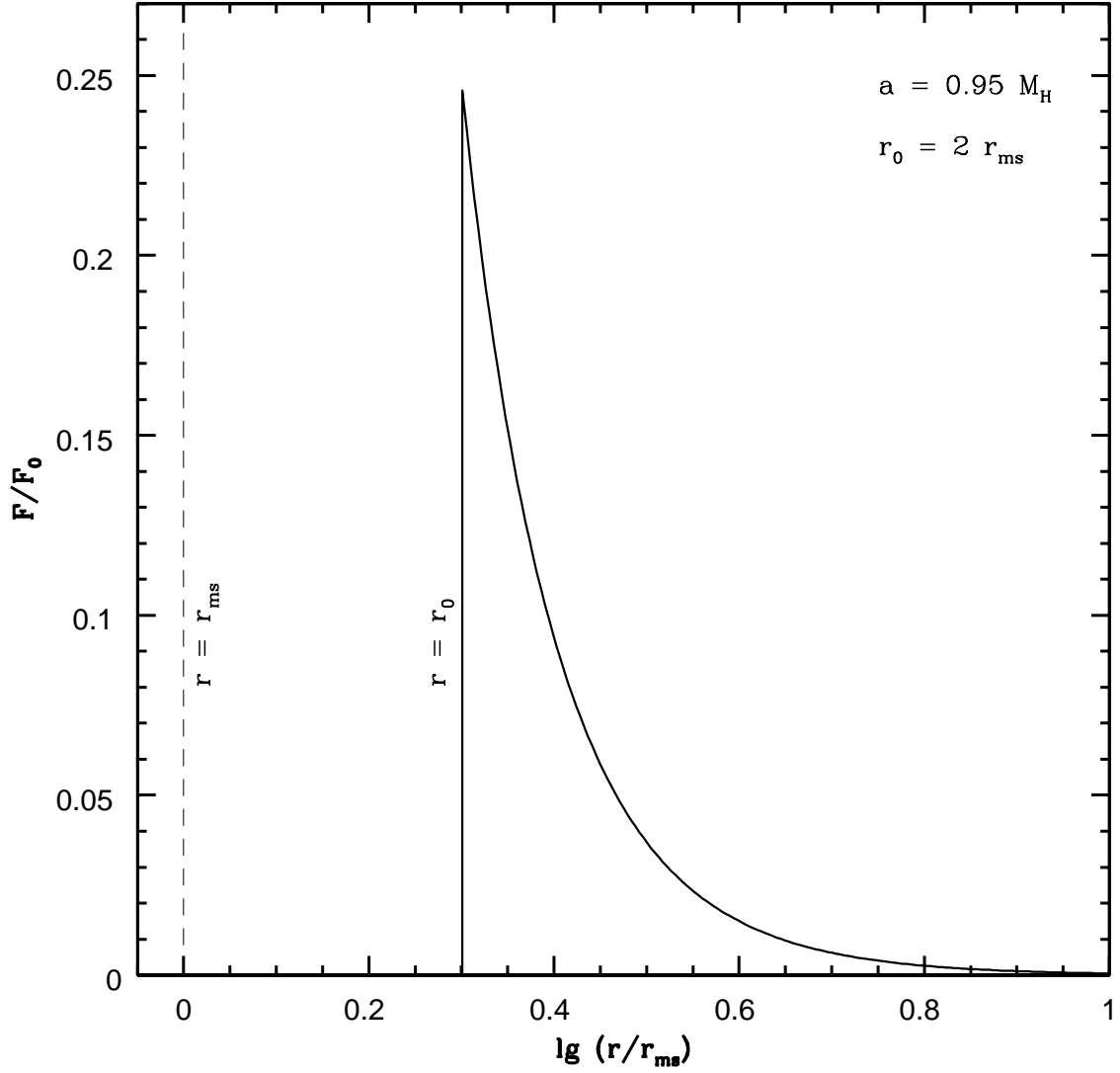}
\caption{The radiation flux of a thin Keplerian non-accretion disk coupled to a Kerr 
black hole with a magnetic field: the magnetic field touches the disk at a circle of 
radius $r=r_0>r_{ms}$, the black hole rotates faster than the disk (i.e. $\Omega_H > 
\Omega_0$). The radius is in unit of $r_{ms}$ -- the radius of the marginally stable
orbit, the radiation flux is in unit of $F_0 \equiv A_0 /r_{ms}^2$, where $A_0$ is 
given by equation (\ref{h0}). The inner boundary of the disk is at $r=r_{ms}$, as 
indicated by the vertical dashed line. The radiation flux is shown with the solid 
curve. For $r<r_0$ (i.e. the left side of the vertical solid line), the radiation 
flux is zero. The radiation flux has a sharp peak at $r = r_0$, and decreases quickly 
for $r>r_0$. The radiation flux approaches $F\propto r^{-3.5}$ at $r\gg r_0$.
\label{fig2}}
\end{figure}

\begin{figure}
\epsscale{0.9}
\plotone{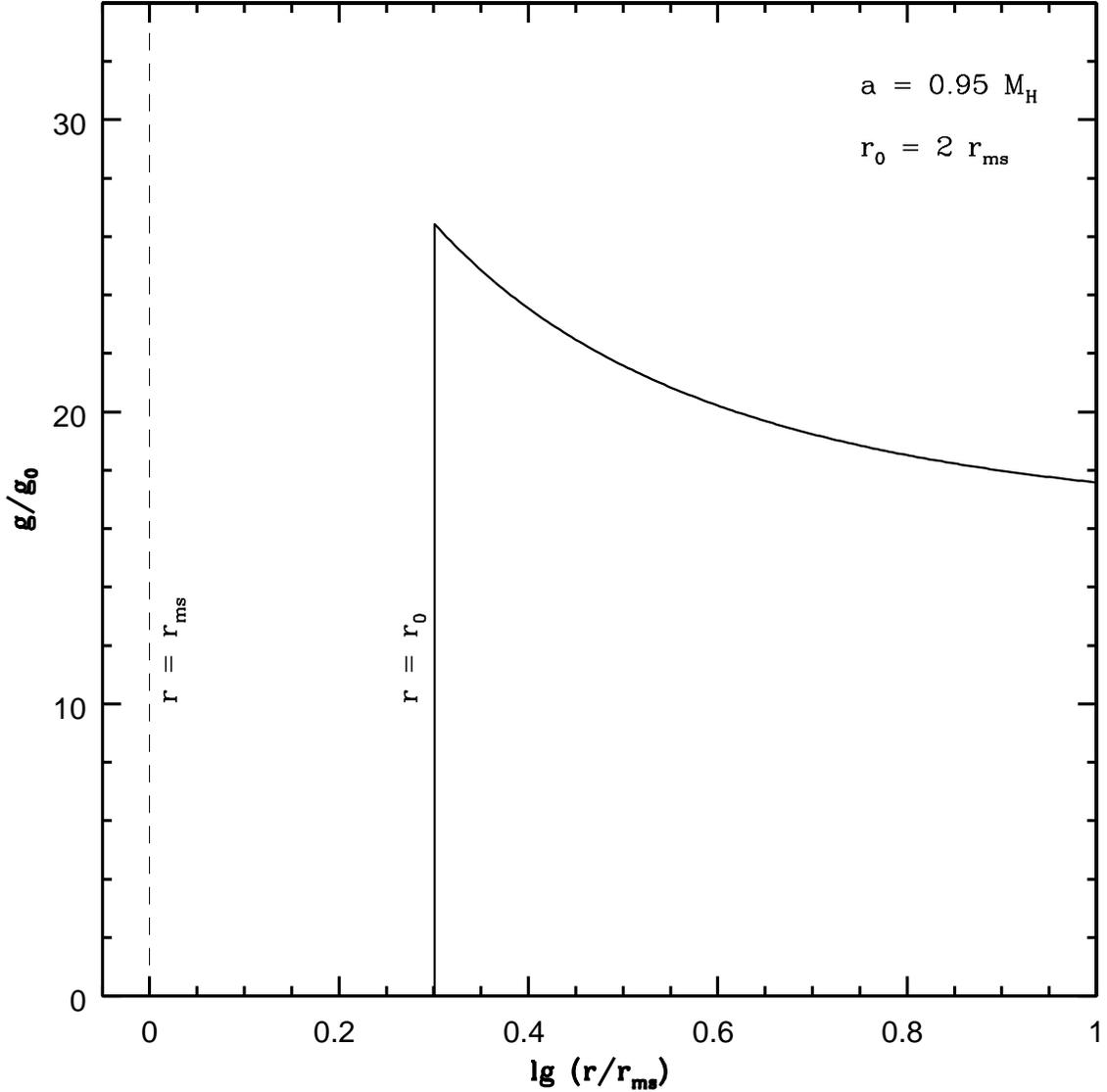}
\caption{The internal viscous torque of a thin Keplerian non-accretion disk coupled
to a Kerr black hole with a magnetic field. The model is the same as that in Fig. 
\ref{fig2}: there is no accretion, the magnetic field touches the disk at a circle 
of radius $r=r_0>r_{ms}$, and the black hole rotates faster than the disk. The radius 
is in unit of $r_{ms}$, the torque is in unit of $g_0 \equiv A_0 r_{ms}$. The inner
boundary of the disk is at $r=r_{ms}$, as indicated with the vertical dashed line. 
The internal viscous torque is shown with the solid curve. For $r<r_0$ (i.e. the left
side of the vertical solid line), the torque in the disk is zero. (Especially, the
torque is zero at $r=r_{ms}$, which is true even for the case of $r_0$ approaching
$r_{ms}$.) The torque rises suddenly at $r=r_0$, then decreases and approaches a
constant at $r\gg r_0$.
\label{fig3}}
\end{figure}

Substituting $\dot{M}_D =0$ into equation (\ref{fdel}) and equation (\ref{gdel}) 
and using the expressions for $E^+ -\Omega_D L^+$ and $d\Omega_D/dr$ given by 
\citet{pag74}, we obtain\footnote{A similar formula for $F$ was obtained by Agol
and Krolik (2000), who treated a non-accretion disk magnetically connected to a
black hole as the ``infinite efficiency limit'' of an accretion disk magnetically
connected to the material in the transition region. However, we emphasize that in 
our model of a disk magnetically connected to a black hole the torque and the 
radiation flux of the disk are always zero at the inner boundary of the disk.
This fact is manifested by the step functions in eq. (\ref{fdel1}) and eq. 
(\ref{gdel1}), which are absent in Agol and Krolik's formula. Thus, as we have 
pointed out in the Introduction, the two models are distinctly different from 
each other.}
\begin{eqnarray}
    F = {3 T_0\over 8\pi}\,\left({M_H\over r^7}\right)^{1/2}\,
        {1\over 1-{3M\over r} + 2a \left({M\over r^3}\right)^{1/2}}\,
        \vartheta\left(r-r_0\right)\,,
        \label{fdel1}
\end{eqnarray}
and
\begin{eqnarray}
    g = {T_0\left[1+ a\left({M\over r^3}\right)^{1/2}\right]\over
         \left[1-{3M\over r}+ 2a \left({M\over r^3}\right)^{1/2}
         \right]^{1/2}}\,
         \vartheta\left(r-r_0\right)\,,
        \label{gdel1}
\end{eqnarray}
where $\vartheta$ is the step function
\begin{eqnarray}
    \vartheta\left(r-r_0\right) = \left\{\begin{array}{ll}
       1, & \mbox{if $r>r_0$}\\
       0, & \mbox{if $r<r_0$}
       \end{array}
       \right.\,,
\end{eqnarray}
and
\begin{eqnarray}
    T_0 &\equiv& 4\pi r_0 A_0 \left(E_0^+ - \Omega_0 L_0^+\right)
                 \nonumber\\
        &=& 4\pi r_0 A_0 \left[1+ a\left({M\over r_0^3}\right)^{1/2}\right]^{-1}
            \left[1-{3M\over r_0}+ 2a \left({M\over r_0^3}\right)^{1/2}
            \right]^{1/2}\,.
\end{eqnarray}
Since $F$ and $g$ cannot be negative, such non-accretion solutions can exist only 
if $A_0 >0$, which requires $\Omega_H> \Omega_0$, i.e. the black hole rotates faster 
than the disk at $r=r_0$. Examples of $F$ and $g$ are plotted in Fig. \ref{fig2} 
and Fig. \ref{fig3}. These figures clearly show that the magnetic coupling takes 
effect only beyond the radius $r=r_0$, i.e. the torque produced by the magnetic
coupling propagates only outward from $r = r_0$. Both $F$ and $g$ have sharp peaks 
at $r=r_0$, are zero for $r<r_0$. Since $r_{ms} < r_0$, the torque and the radiation
flux are always zero at the inner boundary of the disk. At $r=r_0$, the radiation 
flux rises suddenly from zero to a sharp peak, then decreases quickly for $r>r_0$. 
At $r\gg r_0$, the radiation flux decreases with radius as $F\propto r^{-3.5}$. At 
$r=r_0$, the internal viscous torque rises suddenly from zero to a finite value,
then decreases slowly for $r>r_0$ and approaches a constant at $r\gg r_0$.

The radial profile of the radiation flux produced by the magnetic coupling is very
different from that produced by accretion. For a standard accretion disk, the 
radiation flux is zero at $r=r_{ms}$ (the inner boundary of the disk), gradually 
rises to a maximum at a radius beyond $r_{ms}$, then decreases slowly, and approaches 
$F\propto r^{-3}$ at large radii\footnote{At large radii the internal torque 
approaches $g\propto r^{1/2}$ for a standard thin accretion disk.} 
\citep{nov73,pag74,tho74}. While for a non-accretion disk magnetically coupled to a
Kerr black hole, assume the magnetic field touches the disk at the inner boundary, 
then at $r=r_{ms}$ the radiation flux suddenly rises from zero to a sharp peak, then
decreases quickly and approaches $F\propto r^{-3.5}$ at large radii. To compare the 
radiation profile of the
magnetic coupling with the radiation profile of accretion, in Fig. \ref{fig4} we
plot both the radiation flux of a non-accretion disk magnetically coupled to a
rapidly rotating black hole and the radiation flux of a standard accretion disk
rotating around the same black hole. For the non-accretion disk, the magnetic field 
is assumed to touch the disk at the inner boundary (the marginally stable orbit). 
Obviously, the radiation flux of the non-accretion disk has a much steeper radial
profile and a sharp peak closer to the center of the disk, compared to the radiation
flux of the standard accretion disk. For the same models, in Fig. \ref{fig4a} we 
show the emissivity index defined by $\alpha\equiv - d\ln F/d\ln r$, which measures 
the slope of the radial emissivity profile in the disk. We see that, throughout the 
disk the emissivity index for the non-accretion disk with magnetic coupling is 
significantly bigger than the emissivity index for the standard accretion disk. At
large radii, $\alpha$ approaches $3.5$ for the non-accretion disk, $3$ for the standard
accretion disk.

Inserting equation (\ref{hdel}) into equation (\ref{flux_ang}) and equation
(\ref{phd}), we get
\begin{eqnarray}
    T_{HD} = 2\pi A_0 r_0\,, \hspace{1cm} P_{HD} = 2\pi A_0 \Omega_0 r_0\,,
    \label{tphd}
\end{eqnarray}
where $A_0$ is given by equation (\ref{h0}). As expected, $P_{HD} = T_{HD}
\Omega_0$. Since a disk has two surfaces, the total power of the disk is ${\cal 
L}_{HD} = 2 P_{HD} = 4\pi A_0 \Omega_0 r_0\,$. The energy radiated per unit time
from the region inside a circle of radius $r>r_0$ in the disk is
\begin{eqnarray}
    {\cal L}_{HD}(<r) &=& 2 \int_{r_{ms}}^r E^+ F 2\pi r dr \nonumber\\
                     &=& 4\pi \int_{r_{ms}}^r H \Omega_D r dr - g \Omega_D 
                         \nonumber\\
                     &=& 4\pi A_0\Omega_0 r_0 \left(1-{\Omega_D\over\Omega_0}\,
                         {E_0^+ -\Omega_0L_0^+ \over E^+ -\Omega_D L^+}\right)\,,
    \label{mcL12}
\end{eqnarray}
where on the right hand side in the second line we have used equation (\ref{con_ener}) 
(taking $\dot{M}_D = 0$) and the boundary condition $g\Omega (r=r_{ms}) = 0$, in  the 
third line we have used equation (\ref{hdel}) and equation (\ref{gdel}) (taking 
$\dot{M}_D = 0$). We can define a half-light radius $r_{1/2}$, within which the energy 
radiated per unit time by the disk is one half of the total power of the disk, i.e. 
${\cal L}_{HD}(<r_{1/2}) = {1\over 2} {\cal L}_{HD}$. From equation (\ref{mcL12}), for 
a non-accretion disk magnetically coupled to a black hole, $r_{1/2}$ can be solved 
from
\begin{eqnarray}
    \left.{E^+ - \Omega_D L^+ \over \Omega_D}\right\vert_{r=r_{1/2}} = 
        {2\left(E_0^+ - \Omega_0 L_0^+\right) \over \Omega_0}\,.
    \label{r12_mc}
\end{eqnarray}  
Similarly, for a standard accretion disk, the energy radiated per unit time from the 
region inside a circle of radius $r>r_{ms}$ in the disk is
\begin{eqnarray}
    {\cal L}_{acc}(<r) = \dot{M}_D \left(E^+ - E_{ms}^+ - {E^+ -\Omega_D L^+
        \over -d\Omega_D/dr} \Omega_D f\right)\,.
    \label{accL12}
\end{eqnarray}
The total power of an accretion disk is ${\cal L}_{acc} = \dot{M}_D \left(1-
E_{ms}^+\right)\,$. Thus, the half-light radius of a standard accretion disk, which is 
defined by ${\cal L}_{acc}(<r_{1/2}) = {1\over 2} {\cal L}_{acc}$, can be solved from
\begin{eqnarray}
    \left[E^+ - {E^+ -\Omega_D L^+ \over -d\Omega_D/dr} \Omega_D
         f\right]_{r=r_{1/2}} = {1\over 2} \left(1+
         E_{ms}^+\right)\,,
    \label{r12_acc}
\end{eqnarray}
where $f$ is give by equation (15n) of \citet{pag74}.

We have calculated the half-light radius of a disk magnetically coupled to a Kerr 
black hole, assuming the disk has no accretion and the magnetic field touches the disk 
at the inner boundary. The results are shown in Fig. \ref{fig5}. For comparison, we 
have also calculated the half-light radius of a standard accretion disk around a Kerr 
black hole, the results are also shown in Fig. \ref{fig5} with the dashed curve. From 
these results we see that, for a non-accretion disk magnetically coupled to a Kerr 
black hole with the magnetic field touching the disk at the inner boundary, most energy 
radiated by the disk comes from a region closer to the center of the disk, compared to 
the case of a standard accretion disk. A similar figure is shown by Agol and Krolik 
(2000, Fig. 1), and very similar results are obtained by them for a disk magnetically 
coupled to the material in the transition region. But we emphasize that in their model 
a state with a zero accretion rate and a finite power can never be realized, since in 
their model in order to extract energy from a black hole material with negative energy 
must fall into the black hole. And, in our figure the curve is broken at $a/M_H = 
0.3594$ for the non-accretion disk, while in Agol and Krolik's figure the curve is 
drawn without broken.

Suppose a Kerr black hole loses its energy and angular momentum through the magnetic 
coupling to a thin Keplerian disk with no accretion, with the magnetic field lines 
touching the disk at a circle of radius $r=r_0$. Then, the evolution of the black hole 
spin $s = a/M_H = J_H/M_H^2$ is given by
\begin{eqnarray}
    {ds\over d\ln M_H} = {1\over \omega} - 2 s\,,
    \label{dsm}
\end{eqnarray}
where $\omega$ is defined by equation (\ref{omega}), which is a function of $s$ and 
$r_0/r_{ms}$ (see the discussions below eq. [\ref{crit}]). If we know how $r_0/r_{ms}$ 
evolves with $s$, equation (\ref{dsm}) can be integrated to obtain $M_H = M_H(s)$. As 
an example, let us consider a Kerr black hole of initial mass $M_{H,0}$ and initial 
spin $s = 0.998$ -- the spin of a canonical black hole \citep{tho74}. As the black 
hole spins down to $s = s_c$ -- the 
value when $\Omega_H = \Omega_0$ and thus the transfer of energy and angular 
momentum between the black hole and the disk stops, which is defined by equation 
(\ref{crit}), the total amount of energy extracted from the black hole by the disk, 
$\Delta M_H = M_H (s=0.998) - M_H (s=s_c)$,  can be calculated by integrating 
equation (\ref{dsm}). Then we can calculate the fraction of energy that can be 
extracted from the black hole
\begin{eqnarray}
    \eta \equiv {\Delta M_H \over M_{H,0}}
         = 1 - \exp\int_{s_0}^{s_c} {ds\over \omega^{-1} - 2s}\,,
    \label{eta}
\end{eqnarray}
where $s_0 = 0.998$.
For simplicity, we assume $r_0/r_{ms}$ keeps constant as $s$ decreases. The 
corresponding $\eta$ is plotted in Fig. \ref{fig6} as a function of $r_0/r_{ms}$. 
The figure shows that $\eta$ decreases quickly with increasing $r_0/r_{ms}$. The 
maximum fraction is reached when $r_0 = r_{ms}$: $\eta_{max} = 0.152$. Thus, the 
magnetic field lines that touch the inner edge of the disk are most efficient in 
extracting energy from the black hole -- in the sense that the largest amount of 
energy can be extracted from the black hole\footnote{We should note that there are 
two different quantities describing the energetic process for a black hole: the 
fraction of energy extraction, which describes how much energy can be extracted 
from a black hole {\it in total}; the power, which describes how much energy can 
be extracted from a black hole {\it per unit time}. A higher fraction does not 
imply a higher power, and {\it vice versa}. In fact, since $P_{HD}\propto \Omega_D 
\left(\Omega_H - \Omega_D\right)$, $P_{HD}$ peaks when the magnetic field lines 
touch the disk at the place where $\Omega_D = \Omega_H/2$ \citep{li00b}.}.

%\section 4.2
\subsection{$\dot{M}_D \neq 0$
\label{sec4.2}}
In our model there are two torques acting on the disk: one is the external torque
produced by the magnetic coupling to the black hole, the other is the internal torque 
which conveys angular momentum within the disk and dissipates energy. We do not
discuss the origin of the internal torque, we only assume the internal torque exists
and the disk automatically adjusts its internal torque so that quasi-steady state
solutions exist. If the internal torque balances the external torque exactly -- as
assumed in section \ref{sec4.1}, which can be true only if the black hole rotates 
faster than the disk, then a steady state with no accretion is built. In such a case, 
all the power of the disk comes from the rotational energy of the black hole, the 
disk has an infinite efficiency. If the disk has so much internal torque that cannot 
be balanced by the external torque, the excess internal torque will produce
accretion\footnote{Certainly there is yet another possibility: the disk does not
have enough internal torque to remove the angular momentum deposited into the disk
by the black hole, then the accretion flow may be reversed and thus steady state 
solutions do not exist \citep{bla99}.}. In such a case, the power of the disk comes 
from both the rotational energy of the black hole and the gravitational
energy of the disk.

In the case that there is accretion and the magnetic field touches the disk at 
a circle of radius $r_0$, the quasi-steady solutions are given by equation
(\ref{fdel}) and equation (\ref{gdel}). The radiation flux is plotted in Fig. 
\ref{fig7} and Fig. \ref{fig8}, respectively for the case that the black hole 
rotates faster than the disk ($\Omega_H>\Omega_0$) and for the case that the black 
hole rotates slower than the disk ($\Omega_H<\Omega_0$). In both cases, for $r<r_0$ 
the solutions are the same as those predicted by the standard theory of a thin 
Keplerian disk \citep{nov73,pag74,tho74}, in particular $F$ and $g$ are zero at 
$r=r_{ms}$. The extensions of the standard solutions 
to $r>r_0$ are shown with dashed curves, the positions of $r_0$ are shown with 
vertical dotted lines. Due to the magnetic coupling to the black hole, $F$ and $g$ 
are modified for $r>r_0$ by superposing the contribution of the magnetic coupling 
to the standard solutions. For $r\gg r_0$, the radiation flux given by the standard 
theory decreases as $r^{-3}$, while the radiation flux contributed by the magnetic 
coupling decreases as $r^{-3.5}$. Thus, at large radii the radiation flux $F$ is 
dominated by the contribution of accretion. This is also true for the viscous torque 
in the disk, i.e., at large radii the torque in the disk is dominated by the 
contribution of accretion. When the black hole rotates faster than the disk -- as 
the case shown in  Fig. \ref{fig7},  the black hole pumps energy and angular 
momentum into the disk, the energy is locally dissipated in the disk and eventually
radiated away, a bright annular ``bump'' is produced at $r=r_0$. When the black 
hole rotates slower than the disk -- as the case shown in Fig. \ref{fig8}, the 
black hole extracts energy and angular momentum from the disk, and a dark annular 
``valley'' is produced at $r=r_0$.

When the black hole rotates slower than the disk (i.e., $\Omega_H<\Omega_0$), $A_0$
is negative thus energy and angular momentum are transfered from the disk to the 
black hole. From equation (\ref{fdel}), the minimum radiation flux at $r=r_0$ is
\begin{eqnarray}
    F_{\min} = {1\over 4\pi r_0} \left[\dot{M}_D f(r_0) + 4\pi r_0 A_0
       \left(-{d\Omega_0\over dr_0}\right)
       \left(E_0^+ -\Omega_0 L_0^+\right)^{-1}
       \right]\,,
    \label{flux0}
\end{eqnarray}
which (and the minimum $g$ at $r=r_0$) is non-negative if and only if
\begin{eqnarray}
    4\pi r_0 |A_0| \le \left\vert{d\Omega_0\over d r_0}\right\vert^{-1}
     \left(E_0^+ -\Omega_0 L_0^+\right) \dot{M}_D f(r_0)\,,
    \label{positive}
\end{eqnarray}
where we have used the fact that $d\Omega_D/d r <0$ in the disk. Therefore, when the 
black hole rotates slower than the disk, quasi-steady solutions exist only if the 
condition in equation (\ref{positive}) is satisfied. In particular, since $f(r_{ms}) 
= 0$, no quasi-steady solutions exist if the magnetic field touches the disk at the 
inner boundary and the black hole rotates slower than the disk (i.e. $a/M_H<0.3594$).

%\section 5
\section{Weak Magnetic Field Assumption, Instabilities, and Photon Capture}
\label{sec5}
For a non-accretion disk magnetically coupled to a black hole,  a steady state can 
be established if the magnetic
field is frozen in the disk and the angular momentum transfered from the black
hole to the disk is steadily conveyed outward by an internal viscous torque. When
there is accretion, an exactly steady state cannot exist since the magnetic field 
frozen in the disk slowly moves toward the central black hole with the accretion
flow. However, if the inflow velocity of the disk is much smaller than the 
rotational velocity, we can expect the disk and the magnetic field to be in a 
quasi-steady state, which means that any macroscopic quantity at a given radius in 
the disk slowly changes with time: the overall change within one rotation period 
is negligible compared with the quantity itself. For example, the magnetic flux 
within a given radius is $\Psi \sim B r^2$, the change of the magnetic flux within 
one rotation period is $\Delta\Psi \sim B r v_r T \sim B r^2 (v_r / v_\phi)$, where 
$T \approx 2\pi r/v_\phi$ is the rotation period, $v_\phi$ is the rotational 
velocity of the disk, $v_r$ is the radial velocity of the particles in the disk 
(we adopt the convention that $v_r>0$ for accretion). Obviously, we have $\Delta
\Psi \ll \Psi$ for a disk with $v_r \ll v_\phi$. The assumption of a quasi-steady 
state is more or less similar to the assumption of ``adiabatic invariance'' usually
discussed in galactic dynamics and quantum mechanics where it is assumed that a 
potential changes very slowly with time so that within one rotation/oscillation 
period the potential can be treated as unchanged. For a quasi-steady state, the
magnetic connection between the black hole and the disk may evolve with time from 
the view of a long time interval, but within one rotation period the magnetic 
connection is approximately unchanged.

In searching for quasi-steady state solutions we have assumed that the magnetic field 
is weak so that its influence on the dynamics of particles in the disk is negligible,
and thus a thin Keplerian disk is a good approximation. This ``weak magnetic field
assumption'' requires that
\begin{eqnarray}
    \left|\vec{\nabla} \left(B^2/8\pi\right)\right|\ll \rho \left|\vec{g}\right|
    \label{weak1}
\end{eqnarray}
in the disk, where $\rho$ is the mass density of the disk, $\vec{g}$ is the
gravitational acceleration produced by the black hole. Since $\left|\vec{\nabla}
\left(B^2/8\pi\right)\right|\sim B^2 /r$, $\left|\vec{g}\right|\sim GM_H/r^2 \sim
r_H/r^2$, equation (\ref{weak1}) is equivalent to
\begin{eqnarray}
    B^2 \ll \rho c^2\left({r_H\over r}\right)\,,
    \label{weak}
\end{eqnarray}
where we have restored the speed of light in the equation. Equation (\ref{weak}) is
the condition for the weak magnetic field assumption.

A weak magnetic field may play an important role in the balance and transportation 
of energy and angular momentum. From equation (2) of \citet{li00b} and $\Psi_{HD}
\sim Br^2$, we have
\begin{eqnarray}
    {\cal L}_{HD} = 2 P_{HD} \sim B^2 r^2 \left({r\over r_H}\right)^{1/2}
    \label{lhd}
\end{eqnarray}
for the case of $a/M_H\approx 1$, where we have used $\Delta Z_H\sim 1$. The 
accretion power of the disk is
\begin{eqnarray}
    {\cal L}_{acc} = \dot{M}_D \epsilon_0 \sim \rho r h v_r\,,
    \label{lacc}
\end{eqnarray}
where $\epsilon_0$ is the efficiency of accretion ($\approx 0.32$ for $a/M_H = 0.998$,
see eq. [\ref{eff}]), $h$ is the thickness of the disk. From equation (\ref{lhd}) 
and equation (\ref{lacc}), ${\cal L}_{HD} \gtrsim {\cal L}_{acc}$ if and only if
\begin{eqnarray}
    B^2 \gtrsim \rho c^2\left({h\over r}\right)\left({r_H\over r}\right)
    \left({v_r\over v_\phi}\right)\,,
    \label{bal2}
\end{eqnarray}
where we have used $v_r\sim r^{-1/2}$. Equation (\ref{bal2}) is the condition for 
the magnetic field to be important in the balance and transportation of energy and 
angular momentum.

For a thin and quasi-steady accretion disk, we have $h/r\ll 1$ and $v_r/v_\phi\ll 
1$. In the disk we always have $r_H/r\lesssim1$. Therefore, equation (\ref{bal2}) is 
not a too stringent restriction on the values of $B$. To see this, we can compare 
equation (\ref{bal2}) with equation (\ref{weak}). There is a large room for $B^2$ to 
satisfy both equation (\ref{weak}) and equation (\ref{bal2}) if
\begin{eqnarray}
    {v_r\over v_{\phi}}\ll {r\over h}\,.
    \label{room}
\end{eqnarray}
Equation (\ref{room}) is always satisfied by a thin and quasi-steady disk.

In all our analyses the instabilities of the magnetic field and the disk are 
ignored, but in practice they may be important. Because of the outwardly decreasing 
differential rotation, a plasma disk threaded by a weak magnetic field is subject to 
the Balbus-Hawley instability \citep{bal91,bal98}. Within a few rotational periods 
the magnetic field lines in the disk become chaotic and tangled, the disk becomes 
turbulent. This magneto-rotational instability is assumed to play an important role 
in transporting angular momentum within the disk, which is important for producing 
accretion \citep{bal98,haw00a,haw00,men00}. For a thin disk, the effect of magnetic 
reconnection becomes important if the poloidal magnetic field lines threading the 
disk are twisted too much by the rotation of the disk and the central object. This 
may put a restriction on the amplitude of the toroidal magnetic field near the plane 
of the disk, which in turn restricts the power of the energy and the angular 
momentum transfered between the central object and the disk 
\citep{gho78,gho79a,gho79b,liv92,wan95}. But due to the fact that a black hole has 
a large resistance, the situation here may not be so serious as in the case when the 
central object is a star. The screw instability of the magnetic field plays a similar 
role in limiting the amplitude of the toroidal magnetic field and in turn the power 
and torque of the central black hole \citep{gru99,li00a}. Thus, considering the
effect of magnetic reconnection or the screw instability, the actual power and the 
torque of the black hole may be somewhat smaller than those given by equation
(\ref{pow2}) and equation (\ref{toq2}). If the instabilities are strong enough, the 
situation may change so dramatically that quasi-steady state solutions do not exist. 
All these effects of instabilities will be addressed in detail in future.

Not all photons emitted by the disk escape to infinity. Some return to the disk, 
some are captured by the black hole, due to the gravity of the black hole and the 
shape of the disk \citep{tho74,cun76,ago00}. These effects are especially important 
in the inner disk region when $a/M$ is close to $1$. The radiation returning to 
the disk is dissipated and re-radiated, and eventually reaches infinity or falls 
into the black hole. The radiation flux and the overall power of the disk are affected 
by the photons captured by the black hole and the photons returning to the disk. The 
black hole's capture cross-section is greater for photons of negative angular momentum 
(angular momentum opposite to the spin of the black hole) than for photons of 
positive angular momentum \citep{god70,bar73}. Thus, photon capture tends to spin 
down the black hole. Considering the effect of photon capture, \citet{tho74} has 
shown that a black hole will be prevented from spinning up beyond a limiting state 
with $a/M\approx 0.998$ through accretion from a thin Keplerian disk\footnote{However,
the limit $s=0.998$ may be exceeded if a black hole is spun-up by an external torque,
such as in black hole mergers \citep{ago00}, or through accretion from a thick disk
\citep{abr80}.}. For a standard accretion disk without a magnetic coupling, depending 
on the spin of the black hole, among the energy radiated by the disk up to $6\%$ is 
captured by the black hole, up to $28\%$ returns to the disk. For a disk with a 
magnetic coupling, more energy is captured by the black hole or returns to the disk. 
In an extreme case when the magnetic field lines touch the disk at the inner boundary 
and there is no accretion, depending on the spin of the black hole, among the energy 
radiated by the disk up to $15\%$ is captured by the black hole, up to $58\%$ returns 
to the disk, as in the ``infinite efficiency limit'' case of a disk magnetically 
coupled to the material in the transition region \citep{ago00}. Considering the effect 
of radiation capture, the radiation flux at the inner part of the disk is moderately 
modified. However, in the case of a non-accretion disk magnetically coupled to a Kerr
black hole, the returning radiation dominates at large radii since the radiation due 
to the magnetic coupling scales as $r^{-3.5}$ while the returning radiation scales as 
$r^{-3}$ \citep{ago00}. In our calculations the effects of photon capture (either 
captured by the black hole or returning to the disk) are ignored, we hope to consider
them in future.

%\section 6
\section{Conclusions}
\label{sec6}
For an accretion disk magnetically coupled to a Kerr black hole, quasi-steady state
solutions are obtained by assuming the inflow time-scale of particles in the disk
is much longer than the rotational time-scale -- as adopted in the standard theory
of an accretion disk. Though the magnetic field frozen in the disk slowly moves
toward the central black hole as accretion goes on, the inflow velocity of the
magnetic field and the particles is much smaller than the rotation velocity of the
disk. Thus, within one rotation period the magnetic field configuration can
approximately be regarded as unchanged, and the assumption of a quasi-steady state
can be applied. For a general distribution of the magnetic field connecting the
disk to the black hole, the solutions for the radiation flux and the internal
viscous torque are given by equation (\ref{flux1}) and equation (\ref{torque1}),
which are superpositions of the the contribution from accretion and the contribution
from magnetic coupling. From the view of a long period of time (e.g. with many 
rotation periods), the radiation flux and the internal torque of the disk varies
with time (the magnetic connection may even disappear finally), but equation 
(\ref{flux1}) and equation (\ref{torque1}) give the instant values of the radiation
flux and the internal torque of the disk. These general solutions clearly show that
for any distribution of a magnetic field on the disk, the torque produced by the
magnetic coupling propagates outward only, thus the internal torque and the
radiation flux are always zero at the inner boundary of the disk. Even for the 
extreme case that the magnetic field touches the disk at the inner boundary, the 
internal torque $g$ and the radiation flux $F$ of the disk are also zero at the
inner boundary: beyond the marginally stable orbit, as $r$ decreases, $g$ and $F$ 
increase but suddenly drops to zero at $r=r_{ms}$. We emphasize that this feature 
differs from the case of a disk magnetically connected to the material in the 
transition region, where the magnetic stress is demonstrated to be non-zero at the 
inner boundary and extends into the transition region 
\citep{kro99,ago00,haw00,kro01}.

\begin{figure}
\epsscale{0.9}
\plotone{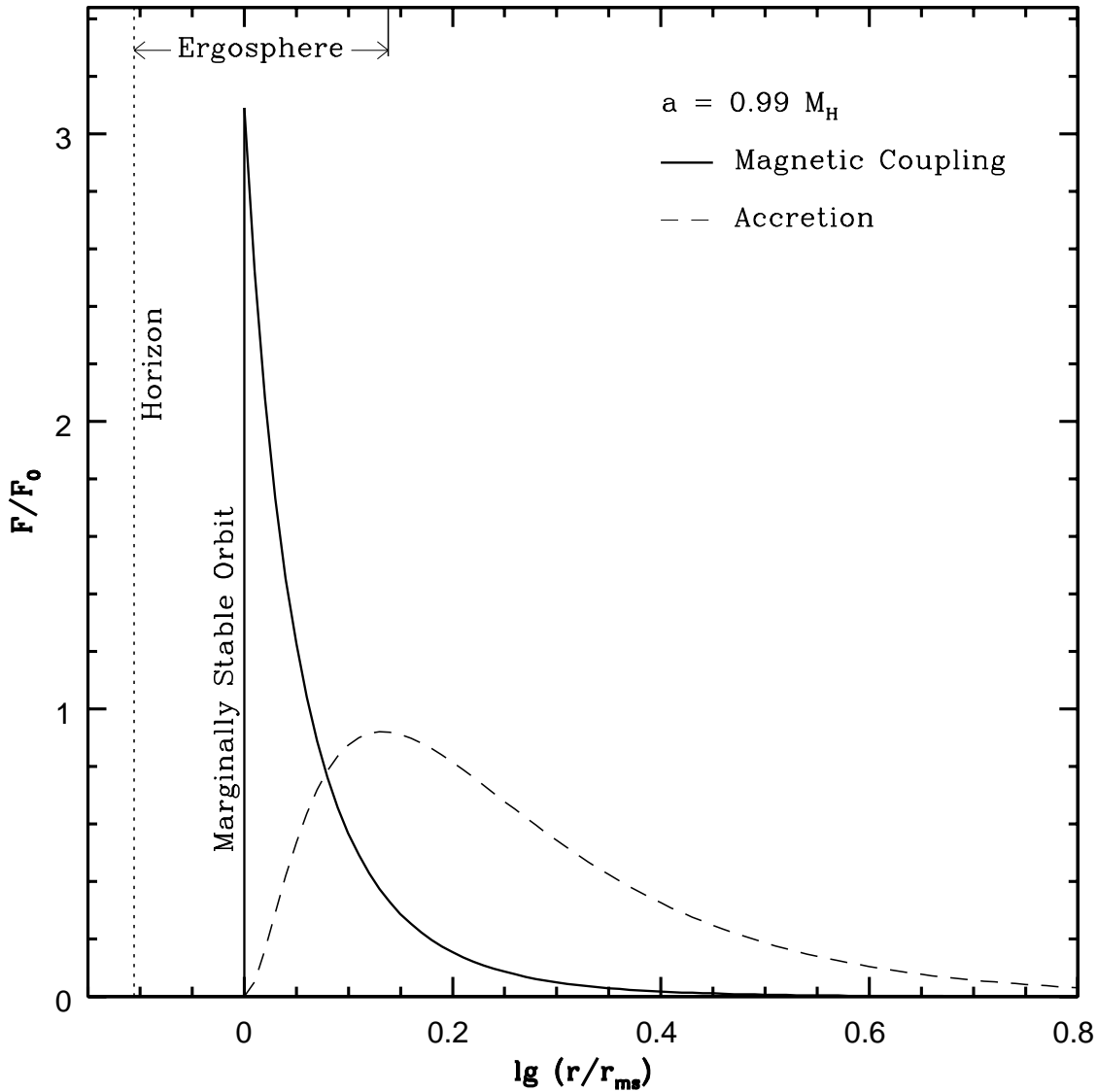}
\caption{Comparison of a non-accretion disk magnetically coupled to a rapidly 
rotating black hole with a standard accretion disk. The horizontal axis is the
logarithm of the radius in the disk, which is in unit of $r_{ms}$ -- the radius
of the marginally stable orbit, which is assumed to be the inner edge of the disk.
The vertical axis is the radiation flux of the disk, which is in unit of $F_0
\equiv A_0/r_{ms}^2$. The solid curve is the radiation flux of a non-accretion disk
magnetically coupled to a Kerr black hole of $a/M_H = 0.99$, where the magnetic
field is assumed to touch the disk at the inner boundary. The dashed curve is the 
radiation flux of a standard accretion disk around a Kerr black hole of $a/M_H = 
0.99$, where the accretion rate is assumed to be $\dot{M}_D = 200 A_0$.
\label{fig4}}
\end{figure}

\begin{figure}
\epsscale{0.9}
\plotone{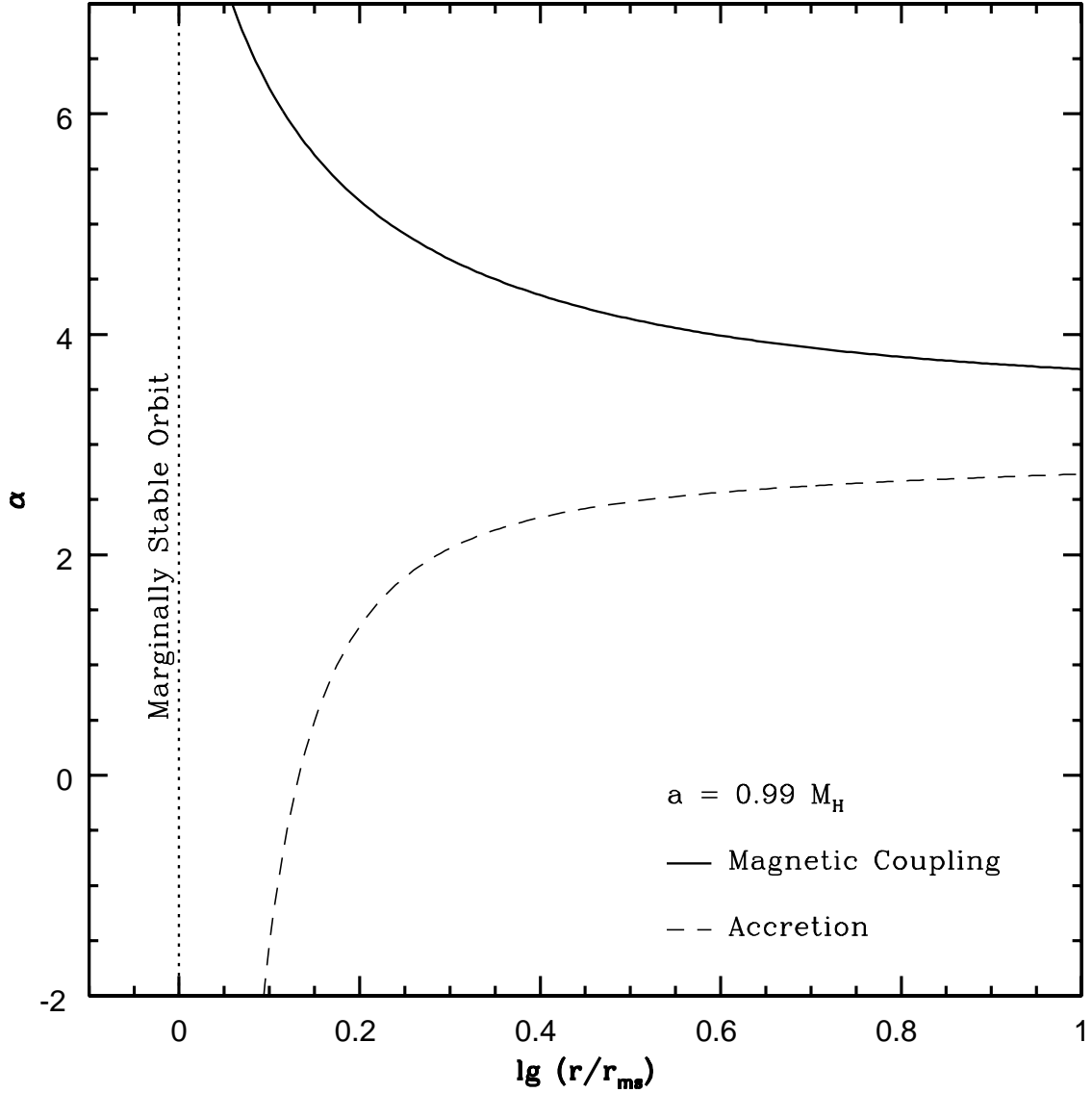}
\caption{The emissivity index: $\alpha\equiv - d\ln F/ d\ln r$. The horizontal axis
is the logarithm of the radius of the disk, which is in unit of $r_{ms}$. The solid 
curve is the emissivity index of a non-accretion disk magnetically coupled to a Kerr 
black hole of $a/M_H = 0.99$, where the magnetic field is assumed to touch the disk 
at the inner boundary. The dashed curve is the emissivity index of a standard 
accretion disk around a Kerr black hole of $a/M_H = 0.99$. Throughout the disk the 
emissivity index for magnetic coupling is significantly bigger than the emissivity 
index for accretion. At large radii, the emissivity index for magnetic coupling 
approaches $3.5$, the emissivity index for accretion approaches $3$.
\label{fig4a}}
\end{figure}

\begin{figure}
\epsscale{0.9}
\plotone{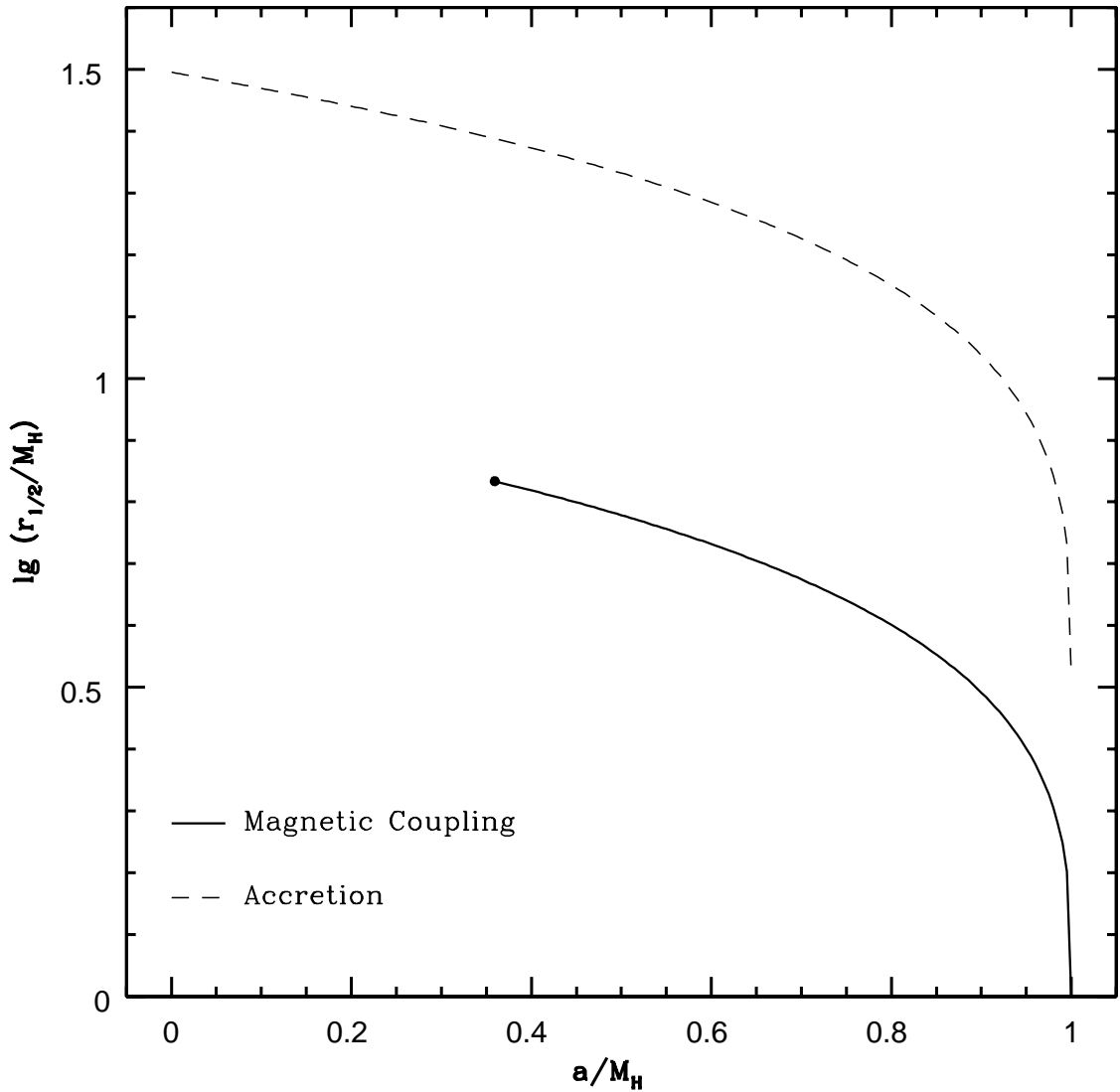}
\caption{The half-light radius of a disk, $r_{1/2}$, as a function of the black hole
spin, $s=a/{M_H}$. The solid curve is the half-light radius of a non-accretion disk
magnetically coupled to a Kerr black hole, where the magnetic field is assumed to
touch the disk at the inner boundary (the marginally stable orbit). The solid curve
starts at $s=0.3594$ (marked by a thick dot), since energy is transfered from the
black hole to the disk only if $s>0.3594$. The dashed curve is the half-light radius
of a standard accretion disk around a Kerr black hole, which exceeds the half-light
radius of a non-accretion disk with magnetic coupling by approximately a half order
of magnitude.
\label{fig5}}
\end{figure}

\begin{figure}
\epsscale{0.9}
\plotone{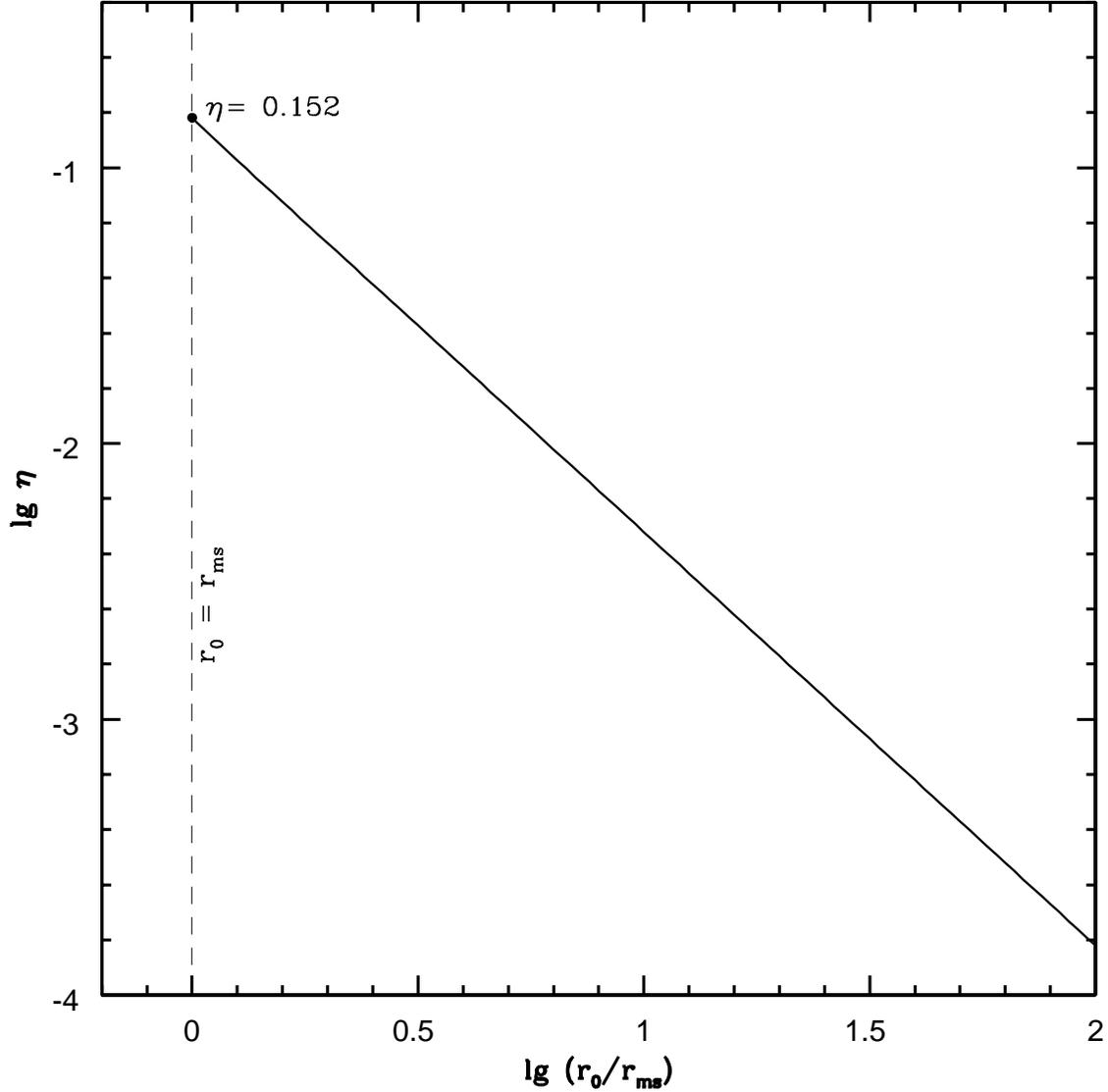}
\caption{The fraction of energy that can be extracted from a Kerr black hole as 
a function of the radius where the magnetic field lines touch the disk. The disk's 
inner boundary is at $r=r_{ms}$ -- the marginally stable orbit as indicated by the 
vertical dashed line. There is no accretion, the magnetic field touches the disk at 
a circle with a radius $r_0> r_{ms}$. The black hole is spun down by the magnetic 
coupling from $s=0.998$, the spin of a canonical black hole, to $s=s_c$, the spin 
at which extraction of energy and angular momentum from the black hole stops. During 
the evolution of the black hole spin the ratio $r_0/r_{ms}$ is assumed to keep 
unchanged. 
\label{fig6}}
\end{figure}

If the black hole rotates faster than the disk, the black hole pumps energy and
angular momentum into the disk through magnetic coupling, which increases the 
efficiency of the disk. Most interestingly, with the existence of the magnetic 
coupling, a disk can radiate without accretion: all the power of the the disk comes 
from the rotational energy of the black hole, such a non-accretion disk has an 
infinite efficiency. We have discussed in detail a simple but important case: the 
magnetic field touches the non-accretion disk at a single radius $r_0$ -- or 
equivalently, the magnetic field is distributed on the disk within a narrow region 
around $r_0$. For this simple case, the radiation flux has a sharp peak at $r_0$: it 
is zero for $r<r_0$ and decreases quickly for $r>r_0$. At large radii, the radiation 
flux approaches $F\propto r^{-3.5}$. This behavior of the radiation flux is very 
different from that of a standard accretion disk, where the radiation flux spreads 
widely over radii and approaches $F\propto r^{-3}$ at large radii. We have compared 
the radiation flux of a non-accretion disk with the magnetic field touching the disk 
at the inner boundary (i.e. the marginally stable orbit), with the radiation flux
of a standard accretion disk in detail. The results are summarized in Fig. 
\ref{fig4} -- Fig. \ref{fig5}. Clearly, the radiation profile of the non-accretion
disk is very different from that of the standard accretion disk: the non-accretion
disk magnetically coupled to a rapidly rotating black hole has a bigger emissivity 
index throughout the disk and a radiation region closer to the center of the disk. 
We have also shown that the magnetic field lines touching the disk at the inner 
boundary is most efficient in extracting energy from the black hole, though the power 
of the black hole peaks at $\Omega_D = \Omega_H /2$.

\begin{figure}
\epsscale{0.9}
\plotone{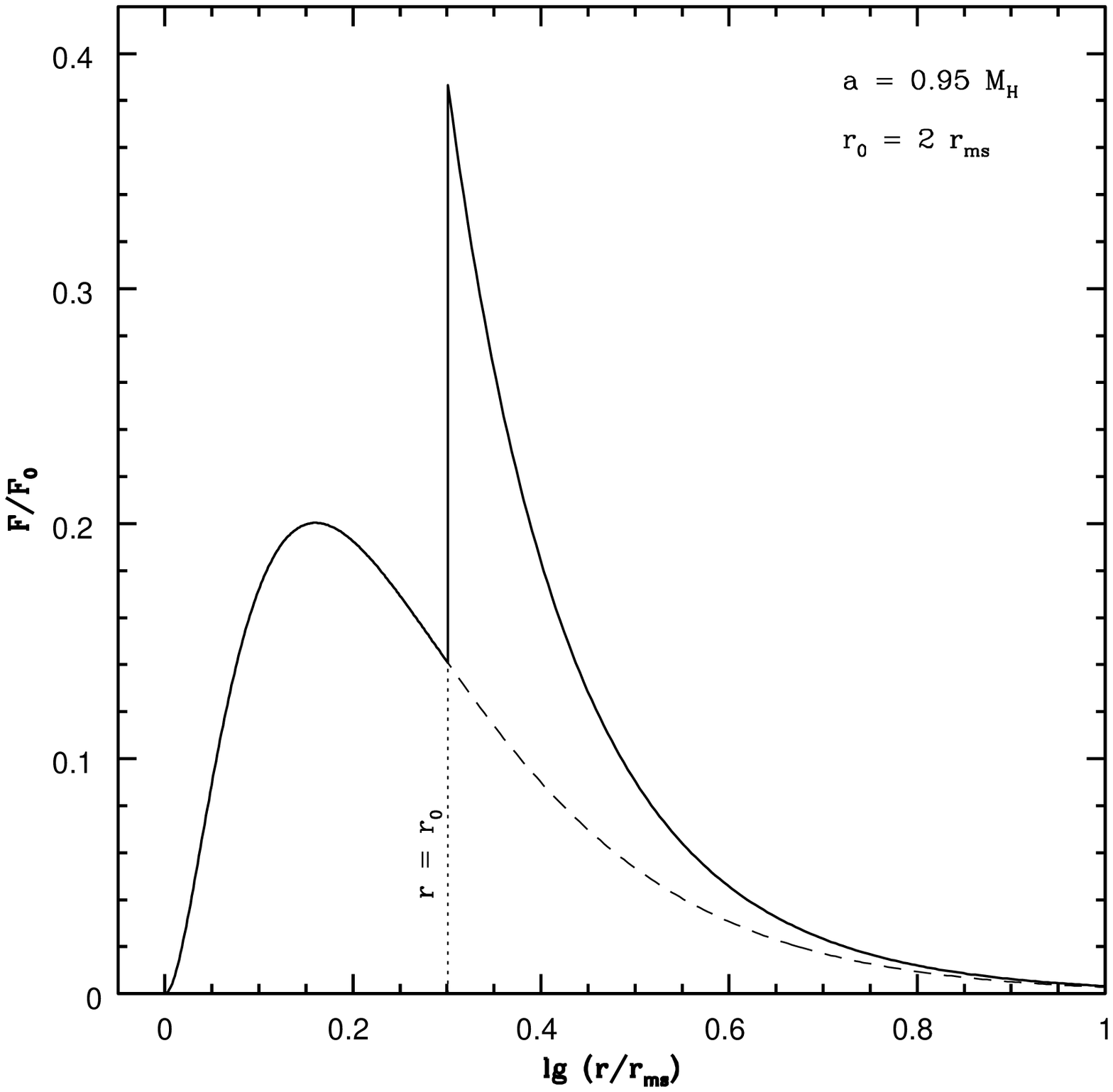}
\caption{The radiation flux of a thin Keplerian accretion disk coupled to a Kerr
black hole with a magnetic field: the black hole rotates faster than the disk. The 
radius is in unit of $r_{ms}$ -- the radius of the marginally stable orbit. The 
radiation flux is in unit of $F_0 \equiv A_0/r_{ms}^2$. The magnetic field
touches the disk at a circle of radius $r=r_0$ which is indicated with the 
vertical solid/dotted line. The solid curve shows the total radiation flux. For 
$r<r_0$ (i.e. the left side of the vertical solid/dashed line), the radiation flux 
is the same as that predicted by the standard theory of an accretion disk, whose 
extension beyond $r=r_0$ is shown with the dashed curve. 
The magnetic coupling produces a bright ``bump'' at $r=r_0$, as indicated by the 
sharp peak in the solid curve. [Parameters for the model: $a/M_H = 0.95$, $r_0/r_{ms}
= 2$, $\dot{M}_D = 80 A_0$.]
\label{fig7}}
\end{figure}

\begin{figure}
\epsscale{0.9}
\plotone{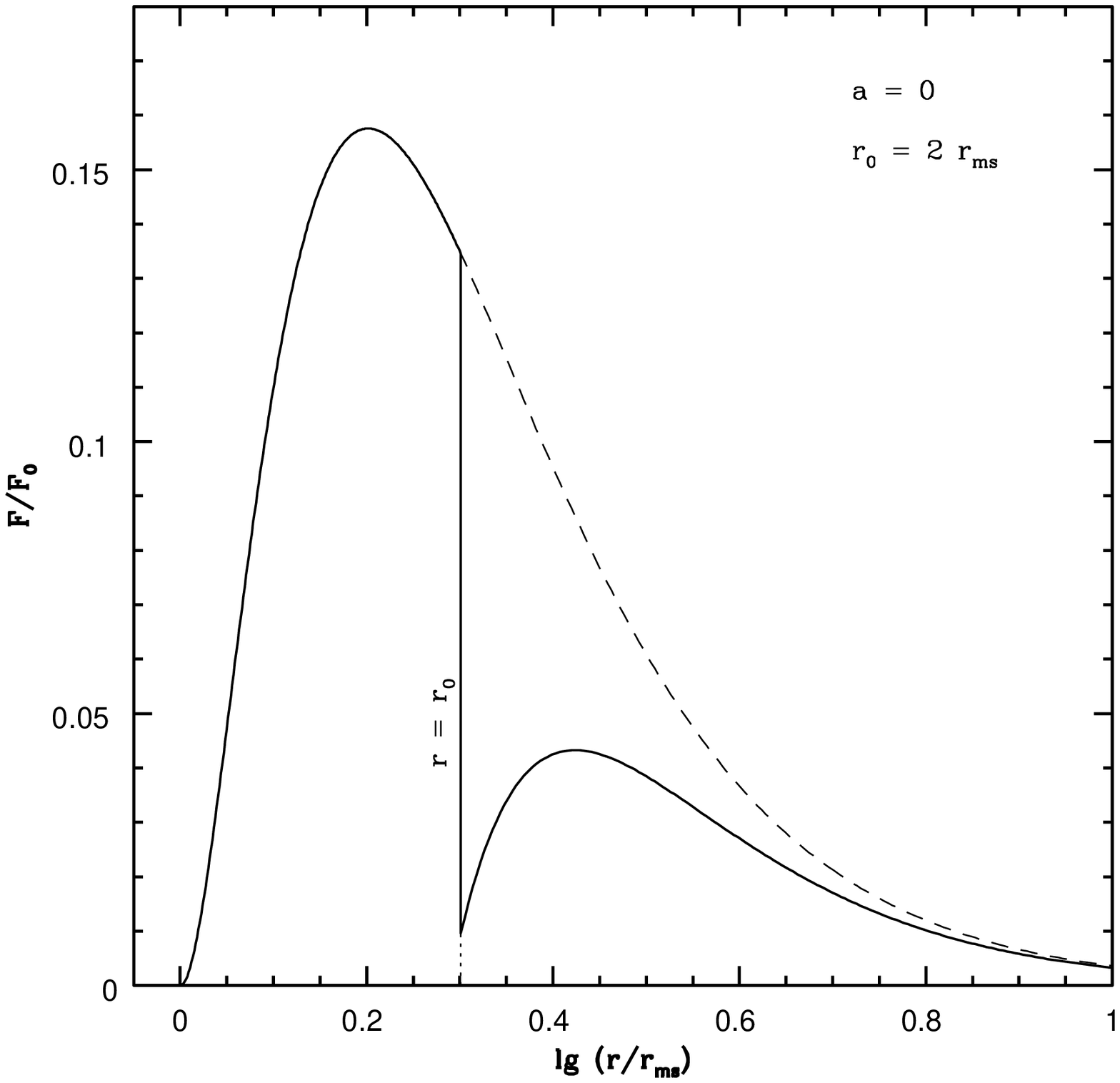}
\caption{The radiation flux of a thin Keplerian accretion disk coupled to a Kerr
black hole with a magnetic field: the black hole rotates slower than the disk. The 
radius is in unit of $r_{ms}$ -- the radius of the marginally stable orbit. The 
radiation flux is in unit of $F_0 \equiv A_0/r_{ms}^2$. The magnetic field
touches the disk at a circle with a radius $r=r_0$ which is indicated with the 
vertical solid/dotted line. The solid curve shows the total radiation flux. For 
$r<r_0$ (i.e. the left side of the vertical solid/dashed line), the radiation flux 
is the same as that predicted by the standard theory of an accretion disk, whose 
extension beyond $r=r_0$ is shown with the dashed curve. The magnetic coupling 
produces a dark ``valley'' at $r=r_0$, as indicated by the deep valley in the solid 
curve. [Parameters for the model: $a/M_H = 0$, $r_0/r_{ms}
= 2$, $\dot{M}_D = 320 A_0$.]
\label{fig8}}
\end{figure}

A non-accretion state can exist only if the internal torque of the disk exactly
balances the external torque on the disk produced by the magnetic coupling. If the 
disk has so much internal torque that cannot be balanced by the external torque, the 
excess internal torque will produce accretion. For an accretion disk with the 
magnetic field touching the disk at a single radius, a specific feature is that the 
radiation flux of the disk may have two maxima (Fig. \ref{fig7} and Fig. \ref{fig8}). 
When the black hole rotates faster than the disk, the black 
hole pumps energy to the disk and the second maximum in the radiation flux is 
produced by adding a bright ``bump'' to the standard radiation flux (Fig. 
\ref{fig7}). When the black hole rotates slower than the disk, the black hole 
extracts energy from the disk -- in such a case the efficiency of the disk is
decreased by the magnetic coupling -- and the second maximum in the radiation flux 
is produced by digging a dark ``valley'' in the standard radiation flux (Fig.
\ref{fig8}). The position of the second maximum is determined by the position where 
the magnetic field touches the disk. If the magnetic field is smoothly distributed 
over the disk instead of touching the disk at a single radius or distributed on the 
disk within a narrow interval of radii, one of the two maxima in the radiation flux 
may be smeared out.

For a black hole with $a/M_H\approx 1$, the power of the black hole is 
approximately
\begin{eqnarray}
    P_{HD}\approx B^2 r_H^2 c\,,
    \label{powhd}
\end{eqnarray}
if the magnetic field lines touch the disk close to the inner boundary ($r_{in} = 
r_{ms}\approx r_H$). If the energy deposited into the disk by the black hole is 
radiated away locally -- either thermally or non-thermally, we can define an 
effective radiation temperature through
\begin{eqnarray}
    T_{eff}\equiv \left({P_{HD}\over\sigma r_H^2}\right)^{1/4}
    \approx \left({B^2 c\over\sigma}\right)^{1/4}
    \approx 5\times 10^5 {\rm K}\,\left({B\over 10^4{\rm Gauss}}\right)^{1/2}\,,
    \label{tem}
\end{eqnarray}
where $\sigma$ is the Stephan-Boltzmann constant. Interestingly, this temperature
depends only on the strength of the magnetic field on the disk. If the energy is 
radiated thermally, the radiation is in the UV to soft X-ray domain for $B\approx 
10^4$ Gauss.

The recent {\it XMM-Newton} observation of soft X-ray emission lines from two
Narrow Line Seyfert 1 galaxies MCG --6-30-15 and Mrk 766 shows an extreme big
emissivity index ($\sim 4$), which has been suggested to indicate that most of 
the line emission originates from the inner part of a relativistic accretion disk 
\citep{bra01}. From the calculations in section \ref{sec4.1}, we have seen that 
for a non-accretion disk magnetically coupled to a rapidly rotating black hole
with the magnetic field touching the disk at the inner boundary, most radiation
comes from a region more concentrated toward the center of the disk and a big
emissivity index can be easily realized at any radius, compared to the case for
a standard accretion disk. Thus, probably the observational results of {\it 
XMM-Newton} can be more easily explained with the model presented in this paper.
Another observation which may be relevant to our model is the kilohertz 
quasi-periodic oscillations (kHz QPOs) in X-ray binaries, which has been 
suggested to originate from the inner edge of a relativistic accretion disk
(van der Klis 2000 and references therein).

The magnetic field connecting the black hole to the disk does not have to be
axisymmetric. If a bunch of magnetic field lines connect the black hole to the 
disk and the feet of the magnetic field lines are concentrated in a small region 
in the disk, a hot spot will be produced on the disk surface if the black hole 
rotates faster than the disk. If $a/M_H\approx 1$, the temperature of the hot 
spot is approximately given by equation (\ref{tem}).

In our analyses the magnetic field has been assumed to be weak and the effect of
instabilities of the disk plasma and the magnetic field has been ignored. The 
situation with a strong magnetic field may be significantly different from that 
has been discussed in this paper, at least the inner part of the disk will not be 
Keplerian any more and quasi-steady state solutions may not exist:  accretion may 
be stopped and a state similar to the ``propeller'' phase for pulsars 
\citep{sch71,ill75,lip92} may be produced, or even the accretion flow is reversed
\citep{bla99}. The instabilities of the disk plasma and the magnetic field may 
seriously affect the dynamics and energetics of the disk, which will make the 
situation much more complicated than that considered in this paper. The effect of 
photon capture caused by the black hole and the disk has also been ignored, which 
is expected to modify both the radiation flux and the total power of the disk. 
All these issues will be addressed in future.

\acknowledgments

I am very grateful to Bohdan Paczy\'nski and Jeremy Goodman for stimulating
discussions and valuable comments. This work was supported by the NASA grant
NAG5-7016 and a Harold W. Dodds Fellowship of Princeton University.

\clearpage

\clearpage

\clearpage

\clearpage

\clearpage

\clearpage

\clearpage

\clearpage

\clearpage

\end{document}